\documentclass[a4paper,11pt]{article}
\pdfoutput=1 
\usepackage{jheppub}

\usepackage{amsmath,graphicx,amssymb,subfigure,bbm}

\usepackage[all]{xy}
\preprint{UTTG-39-13, TCC-004-14}

\title{\boldmath Holographic zero sound at finite temperature\\ in the Sakai-Sugimoto model}

\author[a]{Brandon S. DiNunno,}
\author[b]{Matthias Ihl,}
\author[c]{Niko Jokela,}
\author[a]{and Juan F. Pedraza}
\affiliation[a]{Theory Group, Department of Physics and Texas Cosmology Center,\\
The University of Texas at Austin, Austin, TX, 78712, USA}
\affiliation[b]{Centro de F\'isica do Porto e Departamento de F\'isica e Astronomia,\\
Faculdade de Ci{\^e}ncias da Universidade do Porto,\\ Rua do Campo Alegre 687, 4169-007 Porto, Portugal}
\affiliation[c]{Departamento de F{\'i}sica de Part{\'i}culas e Instituto Galego de F{\'i}sica de Altas Enerx{\'i}as (IGFAE), \\
Universidade de Santiago de Compostela, E-15782 Santiago de Compostela, Spain}
\emailAdd{bsd86@physics.utexas.edu}
\emailAdd{matthias.ihl@fc.up.pt}
\emailAdd{niko.jokela@usc.es}
\emailAdd{jpedraza@physics.utexas.edu}

\abstract{In this paper, we study the fate of the holographic zero sound mode at finite temperature and non-zero baryon density in the deconfined phase of the
Sakai-Sugimoto model of holographic QCD. We establish the existence of such a mode for a wide range of temperatures and investigate the dispersion relation, quasi-normal modes, and spectral functions of the collective excitations in four different regimes, namely, the collisionless quantum, collisionless thermal, and two distinct hydrodynamic regimes. For sufficiently high temperatures, the zero sound completely disappears, and the low energy physics is dominated by an emergent diffusive mode. We compare our findings to Landau-Fermi liquid theory and to other holographic models.}

\begin{document}
\maketitle
\flushbottom

\section{Introduction}
\label{sec:intro}
Understanding the phase structure of QCD at finite temperature and non-zero baryon density is a difficult task and a major focus of current research. At sufficiently large energies, QCD is weakly interacting and perturbation theory is adequate. However, at low energies QCD is strongly interacting and perturbative methods become unreliable, creating a demand for new theoretical tools. The discovery of the gauge/gravity
correspondence \cite{Maldacena:1997re,Gubser:1998bc,Witten:1998qj} has granted us access to the study of a large class of strongly-coupled non-abelian gauge theories.

One of the first and most successful incarnations of holographic QCD is the Sakai-Sugimoto model \cite{Sakai:2004cn, Sakai:2005yt}, which is dual to a $SU(N_c)$ gauge theory with $N_f$ chiral flavors. This model is obtained by considering $N_f$ $D8$-$\overline{D8}$-branes embedded in the background of $N_c$ D4-branes compactified on a circle (with antiperiodic boundary conditions for the fermions). If $N_f\ll N_c$, the backreaction of the flavor branes on the geometry can be sensibly neglected and this, in the field theory, corresponds to working in a ``quenched'' approximation which disregards quark loops.
Although many aspects of the Sakai-Sugimoto model closely resemble real-life QCD, this approach is ultimately troubled by the large-$N_c$ limit inherent in the holographic description. In particular, for large-$N_c$, the low-temperature low-density phase of holographic QCD becomes a crystalline solid instead of the Fermi liquid for the $N_c=3$ case. Moreover, considering the leading $1/N_c$ corrections is a difficult task because it requires doing loop calculations in the bulk.

On the other hand, the deconfined phase of holographic QCD behaves as a diffusive conductor with restored chiral symmetry and may be identified with a strongly coupled liquid \cite{Kim:2008bv}. Indeed, there are some indications that in this regime holographic QCD might be in a Fermi liquid phase. First, the density dependent part of the heat capacity at low temperature is linear in $T$ \cite{Kulaxizi:2008jx}.\footnote{In the Sakai-Sugimoto model, the deconfined phase occurs for $T>T_{\rm cr}$ where $T_{\rm cr}$ is a critical temperature set by the Kaluza-Klein scale. Here, by low temperature we mean $T\ll\mu$, where $\mu$ is the baryonic chemical potential.} This behavior is expected for systems with a Fermi surface, where only a fraction of quasiparticles is excited at small temperatures. Moreover, in the phase where chiral symmetry is restored, the energy density varies as $n_B^{5/3}$, which is the expected power for non-relativistic fermions \cite{Kim:2007vd}.\footnote{In fact, this behavior holds for 
temperatures below $T_{\rm cr}$ and high enough baryon densities \cite{Bergman:2007wp,Rozali:2007rx}. In this regime, the nuclear matter is still in a solid phase and is characterized by a condensate of instantons on the probe flavor branes.}

It is well known that, at sufficiently low temperatures, most metals can be described by Landau-Fermi liquid theory, a phenomenological model where the behavior of interacting fermions are given in terms of quasiparticle excitations of a Fermi surface. In particular, in perturbative QCD the existence of a Fermi liquid phase at high enough chemical potential is well established \cite{Son:2000xc,Son:2000by,Evans:1998ek}, but it is an interesting question whether one can still find signatures of a sharp Fermi surface at strong coupling. This seems unlikely at first sight, but in the large 't Hooft coupling limit, quark-quark interactions are suppressed and there is a chance of observing such behavior at least in some regions of the parameter space. 

One of the main features reminiscent of Fermi liquids is the existence of a gapless excitation in the longitudinal current-current correlators \cite{Landau}. This collective excitation, known as the zero sound, 
arises from oscillations of the Fermi surface which change its shape but not its size. In the context of holography the existence and characterization of such mode has been studied extensively in the literature, starting with the seminal paper \cite{Karch:2008fa}, generalized to non-zero temperature \cite{Bergman:2011rf} and to magnetic field strength \cite{Jokela:2012vn}, with an ever-growing body of work that includes \cite{Kulaxizi:2008kv,Hung:2009qk,Edalati:2010pn,HoyosBadajoz:2010kd,Nickel:2010pr,Lee:2010ez,Ammon:2011hz,
Davison:2011ek,Davison:2011uk,Goykhman:2012vy,Gorsky:2012gi,Brattan:2012nb,Jokela:2012se,Pang:2013ypa,
Dey:2013vja,Edalati:2013tma,Davison:2013uha}. These studies have shed light on the question of whether a true Fermi liquid description can be recovered at low energies in a holographic setup.\footnote{An interesting construction leading to a Fermi liquid behavior was recently proposed in \cite{Rozali:2014yva}. In this paper, the authors considered a phenomenological model in which the bosonic sector is governed by the DBI action, and whose charged sector is purely fermionic. It would be interesting to study the longitudinal current-current correlators for this setup and determine if it supports the existence of a zero sound mode.}$^{,}$\footnote{There has also been a lot of interest in understanding non-Fermi liquids and marginal Fermi liquids in the context of holography \cite{Liu:2009dm, Iqbal:2011ae, Jensen:2011su}. These are situations with non-trivial IR fixed points where a Fermi surface still exists but the quasiparticle description breaks down, and could play an important role in understanding, e.g., 
the strange metal phase of high $T_c$ cuprates and thus high temperature superconductivity. }

In the deconfined phase of holographic QCD, in particular, it was found that a zero sound mode exists at least in the zero temperature limit $T/\mu\ll1$ \cite{Kulaxizi:2008jx}. However, it is well-known that for such high densities the model is unstable towards formation of baryon charge density waves \cite{Ooguri:2010xs}. The reason behind this is that there are Chern-Simons terms for the gauge fields which typically lead to spatially modulated phases, similar to other \#ND=6 cousins, D3/D7' \cite{Bergman:2010gm} and D2/D8' \cite{Jokela:2011eb} brane intersection models. An interesting question we may ask here is whether this mode survives at temperatures and densities that are stable against this striped phase formation, and this paper is devoted to answering this question.\footnote{The spatially modulated instabilities can be mitigated and effectively washed out by turning on a sufficiently large magnetic field and/or a mass for the quarks \cite{Bergman:2011rf, Jokela:2012vn, Jokela:2012se}. The former is 
an 
obvious generalization of the present work, albeit a very involved one due to a new phase opening up in the theory \cite{Lifschytz:2009sz} that leads to very interesting physics \cite{Preis:2010cq}. We will leave this issue for a future work.}$^{,}$\footnote{The confining phase can also be unstable due to spatial inhomogeneities upon introducing chemical potentials \cite{Aharony:2007uu,Bayona:2011ab,BallonBayona:2012wx}.} More specifically, we will investigate the fate of the holographic zero sound mode in the deconfined phase of the Sakai-Sugimoto model in the full space of parameters $T$ and $\mu$. We will characterize the different transitions, starting from the expected quantum regime $T/\mu\ll1$ \cite{Kulaxizi:2008jx} and ending with a high temperature regime $T/\mu\gg1$ where the zero sound mode completely disappears and the physics is dominated by an emergent diffusive mode \cite{Kim:2008bv}.

The paper is organized as follows: in section 2, we briefly review some important features of the Sakai-Sugimoto $D4/D8/\overline{D8}$ model relevant to the investigation of the collective excitations in consideration. In section 3, we derive the equations of motion for the longitudinal fluctuations of the gauge field and we study in detail the asymptotic behavior of the solutions for both the near-boundary and near-horizon regions in the bulk. In section 4, we present our numerical results for the dispersion relation, quasi-normal modes, and spectral functions for the various regions of the parameter space. Finally, in section 5 we conclude with a discussion of the important similarities and differences of the zero sound mode in the Sakai-Sugimoto model compared to other holographic models and Landau-Fermi liquid theory, and comment briefly on possible future directions in the research on holographic quantum liquids.

\section{The Sakai-Sugimoto model at finite temperature}
\subsection{General setup}
Let us consider the near horizon geometry generated by the $N_c$ $D4$-branes
\begin{eqnarray}
ds^2&=&\left(\frac{u}{R}\right)^{\frac{3}{2}} \left(-f(u)dt^2+dx_i^2+dx_4^2\right)+
              \left(\frac{u}{R}\right)^{-\frac{3}{2}}\left(\frac{du^2}{f(u)}+u^2 d\Omega_4^2\right),\nonumber \cr
              e^{\Phi}&=&g_s \left(\frac{u}{R}\right)^{\frac{3}{4}}, \qquad\qquad F_4=dC_3=\frac{2\pi N_c}{\Omega_4}\omega_4 ,
\end{eqnarray}
where $t$, $x^i$, $i=1,2,3,$  and $x^4$ represent the dimensions of the worldvolume of the $D4$-branes, $u$ is the radial (holographic)
coordinate and $d\Omega_4^2$ the metric of the unit four-sphere, while $\Omega_4$ and $\omega_4$ denote the
volume and volume form of the unit four-sphere, respectively. The function $f(u):=1-\left(\frac{u_H}{u}\right)^3$ is the usual emblackening factor, where $u_H$, the location of the horizon, is related to the temperature of the field theory via
\begin{equation}\label{eq:defu}
T_H =\frac{3u_H^{1/2}}{4\pi R^{3/2}}.
\end{equation}
This describes the high temperature, deconfined phase of the Sakai-Sugimoto model \cite{Sakai:2004cn, Sakai:2005yt, Aharony:2006da} in Minkowski signature, appropriate for real time dynamics.\footnote{
In the $D3/D7$ model at finite baryon density, the probe brane embeddings are ``black hole" embeddings that fall into the horizon \cite{Kobayashi:2006sb}. This is the analogue of the high-temperature, deconfined phase we are interested in here. Note, however, that in the deconfined phase of the Sakai-Sugimoto, there exists a region of parameter space where the (``short cusp'') U-shaped embeddings are at least meta-stable, even at non-zero baryon density/chemical potential \cite{Bergman:2007wp}.}$^{,}$\footnote{For some recent criticism on the standard mechanism of the deconfinement/confinent phase transition in the Sakai-Sugimoto model, see \cite{Mandal:2011ws, Mandal:2011uq}.}\\
The radius of curvature $R$ is given by
\begin{equation}
R^3=\pi g_s N_c l_s^3=\pi \lambda \alpha',
\end{equation}
where $\lambda$ is the 't Hooft coupling constant. The scale $R_4= \frac{2}{3} \left(\frac{R^3}{u_{\Lambda}}\right)^{1/2}$ determines the critical temperature $T_{\rm cr}=\frac{1}{2\pi R_4}$ at which a Hawking-Page (confinement/deconfinement) phase transition occurs \cite{Aharony:2006da}.\footnote{The parameter $u_{\Lambda}$ sets the position of the tip of the cigar in the zero, or low temperature, confined phase of the Sakai-Sugimoto model and also determines the critical temperature at which the deconfinement transition happens.} In order to stay within the high temperature regime of the model, we demand $T_H > T_{\rm cr}$, or equivalently $u_H > u_\Lambda$.\\
In the zero, or low temperature, confined phase the $N_f \ll N_c$ probe $D8/\overline{D8}$-branes describe a non-trivial profile in the $x_4(u)$ direction and the two branches merge at a radial position $u_0 \geq u_\Lambda$, which is the location of the tip of the cigar-shaped subspace $\{x_4,u\}$. In the high temperature, deconfined phase that we will be interested in, the  $\{x_4,u\}$ subspace will be cylinder-shaped and the straight embeddings, for which $\partial_u x^4 =0$, will be energetically favored.
\subsection{Chemical potential}
Introducing a non-zero chemical potential amounts to introducing a flux $F_{tu}\neq 0$
along the brane worldvolume \cite{Bergman:2007wp, Rozali:2007rx, Kim:2007zm} (see also \cite{Kim:2006gp,Horigome:2006xu,Sin:2007ze,Yamada:2007ys}). The corresponding DBI action reads
\begin{equation}
S_{D8}=-\mathcal{N_T} \int d u \, u^{\frac{5}{2}} \sqrt{1-(\partial_u A_t)^2},
\end{equation}
where $\mathcal{N}_T= \frac{\mathcal{N}}{T}$ and $\mathcal{N}= \frac{\mu_8}{g_s}\Omega_4 R^{\frac{3}{2}}=\frac{\sqrt{2}}{3} (2\pi)^{-\frac{11}{2}} \frac{N_c N_f}{\sqrt{\lambda}}$.
Here we made a gauge choice to set $A_u=0$ and also rescaled $A_t \rightarrow 2 \pi A_t$.
As usual, the conserved charge associated with $A_t$ reads
\begin{equation}
u^{{5\over 2}} \frac{\partial_u A_t} {\sqrt{1-(\partial_u A_t)^2}}=d.
\end{equation}
It follows that the electric field satisfies
\begin{equation}
\partial_u A_t=\frac{d}{\sqrt{d^2+u^5}},
\end{equation}
from which one can read off the chemical potential $\mu$ as the asymptotic
value of $A_t$,
\begin{equation}\label{eq:mu}
\mu= A_t(u\rightarrow \infty) = \int_{u_H}^{\infty} du \, \partial_u A_t=\frac{d}{3\pi u_H^{3/2}}\;{}_2F_1\left(\frac{1}{2},\frac{3}{10},\frac{13}{10},-\frac{d^2}{u_H^5}\right).
\end{equation}
\section{Zero sound mode in the Sakai-Sugimoto model}
\subsection{DBI action and longitudinal fluctuations}
We want to study the massless excitation coupled to the charge density operator
in the $D4/D8/\overline{D8}$-system. This requires analyzing the linearized
equations of motions that follow from the quadratic action describing the fluctuations of the gauge fields living on the $D8/\overline{D8}$-branes.\footnote{The complete $D8$ action consists of a Dirac-Born-Infeld (DBI) and a Chern-Simons (CS)
term. However, for the longitudinal part of the gauge field fluctuations, it suffices to study the DBI part of the action. Note also that the gauge field and metric perturbations decouple.}
The DBI action for the $D8$-brane reads
\begin{equation}\label{eq:DBI}
S_{\mathrm{DBI, D8}}=-T_8 \int d\Omega_4\int d^4x \int_{u_H}^{u_B} d u \; e^{-\Phi}\sqrt{-\det{\left[G_{mn}+\mathcal{F}_{mn}\right]}},
\end{equation}
where $G_{mn}$ and $\mathcal{F}_{mn}$ are the induced metric and gauge field strength, respectively.\\
For the investigation of the zero sound mode, we want to study small (longitudinal) fluctuations of the gauge field. We are only interested in fluctuations which are independent of the $S^4$ coordinates. Therefore we work in a gauge where $\mathcal{A}_u \equiv 0$ and will set $\mathcal{A}_{y} (x^{\mu},u)= \mathcal{A}_z(x^{\mu},u)= 0$.\footnote{This can be done consistently, since the longitudinal fluctuations decouple from the transversal ones, cf. e.g., \cite{Kim:2008bv}.} \\
Let us introduce the following useful functions: 
\begin{equation}
g(u):=\sqrt{d^2+u^5}, \quad f_1(u) := g(u) f(u), \quad f_2(u) :=\frac{g^3(u)}{u^5}, \quad f_3(u):= \frac{R^3 g(u)}{u^3 f(u)},
\end{equation}
in terms of which the DBI action for the longitudinal modes reduces to
\begin{equation}
S_{\mathrm{DBI, D8}}=-\frac{1}{2}\mathcal{N_T} \int d^4x \int_{u_H}^{u_B} d u \left( f_1(u) \left(\partial_u \mathcal{A}_x \right)^2 - f_2(u) \left(\partial_u \mathcal{A}_t \right)^2 - f_3(u)
\left(\partial_x \mathcal{A}_t -\partial_t \mathcal{A}_x \right)^2\right).
\end{equation}
The gauge field is expanded as follows
\begin{eqnarray}
\mathcal{A}_t (x^{\mu},u)&=& A_t (u) +\int \frac{d^4k}{(2 \pi)^4} e^{i k_{\mu} x^{\mu}} a_t(k_{\mu},u),\cr
\mathcal{A}_x (x^{\mu},u) &=& \int \frac{d^4k}{(2 \pi)^4} e^{i k_{\mu} x^{\mu}} a_x(k_{\mu},u),
\end{eqnarray}
where we choose $k_{\mu}=\left(-\omega, k, 0,0\right)$ and sometimes suppress the dependence of $a_{\mu}$ on $k_{\mu}$. \\
Expanding the DBI action to second order in the fluctuations yields the following equations of motion for the longitudinal modes,
\begin{subequations}\label{eq:longeqs}
\begin{align}
\partial_u \left[f_2(u) \partial_u a_t(k_{\mu},u) \right] - k^2 f_3(u) \left( a_t(k_{\mu},u) + \frac{\omega}{k} a_x(k_{\mu},u) \right) &=0,\\
\partial_u \left[f_1(u) \partial_u a_x(k_{\mu},u) \right] + \omega^2 f_3(u) \left( a_x(k_{\mu},u) + \frac{k}{\omega} a_t(k_{\mu},u) \right) &=0.
\end{align}
\end{subequations}
Let us introduce the gauge invariant quantity
\begin{equation}
E(k_{\mu},u):= k a_t(k_{\mu},u) +\omega a_x(k_{\mu},u),
\end{equation}
and moreover impose Gauss' law, which is a constraint equation following from the equation of motion for $a_u(k, u)$,
\begin{equation}\label{eq:constraint}
k a_x' (u) + \omega G(u) a_t'(u) =0,
\end{equation}
where the prime denotes derivation with respect to $u$ and $G(u):=\frac{f_2(u)}{f_1(u)}=\frac{g^2(u)}{u^5 f(u)}$. 
Using Gauss' law it is possible to express
\begin{eqnarray}
a_t' (u) &=& \frac{k}{k^2 - \omega^2 G(u)} E'(u),\; a_t''(u) = \frac{k}{k^2 - \omega^2 G(u)} E''(u)+ \frac{k \omega^2 G'(u)}{\left(k^2-\omega^2 G(u)\right)^2} E'(u),\\
a_x'(u) &=& \frac{\omega G(u)}{\omega^2 G(u) -k^2}E'(u),\; a_x''(u) = \frac{\omega G(u)}{\omega^2 G(u) -k^2}E''(u) - \frac{\omega k^2 G'(u)}{\left(\omega^2 G(u) -k^2\right)^2} E'(u).
\end{eqnarray}
Any one of the eqs.~(\ref{eq:longeqs}), together with the constraint eq.~(\ref{eq:constraint}), implies the remaining one. Defining $F(u):=k^2-\omega^2 G(u)$, we thus arrive at a single second order equation for $E(u)$ (similar equations, albeit in a different context, were discussed in \cite{Parnachev:2006ev}),
\begin{equation}\label{eq:Eueqn0}
E''(u)+\left( \frac{f_2'(u)}{f_2(u)}-\frac{F'(u)}{F(u)}\right) E'(u) - \frac{f_3(u) F(u)}{f_2(u)} E(u) =0,
\end{equation}
or equivalently,
\begin{equation}\label{eq:Eueqn}
E''(u) + \left(- \frac{5}{u}+\frac{15 u^4}{2g^2(u)}+\frac{\omega^2 G'(u)}{k^2-\omega^2 G(u)}\right) E'(u) - \left( \frac{R^3 u^2 \left(k^2 -\omega^2 G(u) \right)}{g^2(u) f(u)}\right) E(u) =0.
\end{equation}
It turns out to be convenient to perform a change of variables
\begin{equation}\label{eq:defy}
y= 2 \sqrt{\frac{R^3}{u}},
\end{equation}
and define $\widehat{\mu}:=\frac{d^{1/5}}{2 R^{3/2}}$,\footnote{Note that the critical density, above which an instability occurs in the transversal sector coupled to the Chern-Simons term, was determined in \cite{Ooguri:2010xs}, eq. (18). For the Chern-Simons coupling in the Sakai-Sugimoto model, this yields
\begin{equation}
\widehat{\mu}_{\mathrm{crit}}\approx 3.714 \frac{2 \pi}{3}T_H \quad \Rightarrow \; \widetilde{\mu}_{\mathrm{crit.}} \approx 3.714.
\end{equation} } which can be expressed in terms of $\mu$ taking into account (\ref{eq:mu}).

In order to facilitate the translation between the $y$ and $u$ coordinates, we define functions similar to the $f_i$ above (by a slight abuse of notation, we will continue to use $f(y)=1-\left(\frac{y}{y_H}\right)^6$ and $g(y) = \sqrt{1+\widehat{\mu}^{10}y^{10}}$),
\begin{equation}
h_1 = \frac{f(y) g(y)}{y^2}, \quad h_2 = \frac{g^3(y)}{y^2}, \quad h_3 = \frac{g(y)}{y^2 f(y)},
\end{equation}
and note the following transformation rules:
\begin{eqnarray*}
u &\leftrightarrow& y,\\
f_1 &\leftrightarrow& h_1, \\
f_2 &\leftrightarrow& h_2, \\
f_3 &\leftrightarrow& h_3, \\
\end{eqnarray*}
Moreover, in order to render eq.~\eqref{eq:Eueqn} dimensionless, we perform the following rescalings,
\begin{equation}
y \rightarrow y_H \, \widetilde{y}, \quad  \omega \rightarrow \frac{\widetilde{\omega}}{y_H}, \quad k \rightarrow \frac{\widetilde{k}}{y_H}, \quad \widehat{\mu} \rightarrow \frac{\widetilde{\mu}}{y_H},
\end{equation}
so that\footnote{For the presentation of the numerical computations reported in section 4, it is sometimes advantageous to work with the alternative dimensionless variables
\begin{equation}\label{eq:barvar}
\overline{\omega}= \frac{\omega}{\hat{\mu}},\qquad \overline{k}= \frac{k}{\hat{\mu}}.
\end{equation}}
\begin{equation}
\tilde{\omega} = \frac{3}{2 \pi} \frac{\omega}{T_H}, \quad \tilde{k} = \frac{3}{2 \pi} \frac{k}{T_H}, \quad \tilde{\mu} = \frac{3}{2 \pi} \frac{\hat{\mu}}{T_H}.
\end{equation}
Eq. \eqref{eq:Eueqn0} then becomes (in dimensionless variables)
\begin{equation}\label{eq:Ezeqn}
\ddot{E}(\widetilde{y}) + \left(\frac{\dot{h}_2(\widetilde{y})}{h_2(\widetilde{y})}-\frac{\dot{F}(\widetilde{y})}{F(\widetilde{y})}\right)\dot{E}(\widetilde{y}) - \frac{h_3(\widetilde{y})F(\widetilde{y})}{h_2(\widetilde{y})} E(\widetilde{y}) = 0.
\end{equation}
Here the dot $\dot{}$ indicates derivation with respect to $\widetilde{y}$. 
\subsection{Asymptotic solution}
We proceed to solve eq.~(\ref{eq:Ezeqn}) by applying a standard Frobenius series expansion.
First note that  eq.~(\ref{eq:Ezeqn}) has non-essential (regular) singular points at $\widetilde{y}=-1,0,+1,\infty, $ and at the real roots of $ \widetilde{y}^6 \left( \widetilde{k}^2+ \widetilde{\omega}^2 \, \widetilde{\mu}^{10}\, \widetilde{y}^{4}\right)=\widetilde{k}^2 - \widetilde{\omega}^2 $. The characteristic exponents at the horizon ($\widetilde{y}=1$) are $\pm \frac{i \tilde{\omega}}{6}$, corresponding to solutions near the boundary ($\widetilde{y}=0$) with incoming-wave ($- \frac{i \tilde{\omega}}{6}$) and outgoing-wave ($+ \frac{i \tilde{\omega}}{6}$) boundary conditions, respectively.
On general grounds, the solution to eq.~(\ref{eq:Ezeqn}) that satisfies an incoming-wave boundary condition near the horizon can be written as a linear combination of two local solutions near the boundary (with exponents 0 and 3) \cite{Son:2002sd, Herzog:2002pc, CaronHuot:2006te}:
\begin{equation}\label{eq:Ebdy}
E(\widetilde{y})= \mathcal{A} Z^I(\widetilde{y}) + \mathcal{B} Z^{II}(\widetilde{y}),
\end{equation}
where
\begin{eqnarray}
Z^I(\widetilde{y})&=& 1 + b^I_1 \widetilde{y} + b^I_2 \widetilde{y}^2 + \ldots, \\
Z^{II}(\widetilde{y})&=& \widetilde{y}^3 \left( 1 +  b^{II}_1 \widetilde{y} + b^{II}_2 \widetilde{y}^2 + \ldots \right).
\end{eqnarray}
The coefficients $b^{I,II}_i$ are determined by recursion relations following from eq.~(\ref{eq:Ezeqn}). For the case at hand, we find 
\begin{eqnarray}
b^I_{1,3,5,\ldots}=0,&~&\qquad b^{II}_{1,3,5,\ldots}=0,\cr
b^I_2 = -\frac{1}{2}\left(\widetilde{k}^2-\widetilde{\omega}^2\right),&~& \qquad b^{II}_2= \frac{1}{10}\left(\widetilde{k}^2-\widetilde{\omega}^2\right), \cr
b^I_4 = -\frac{1}{8}\left(\widetilde{k}^2-\widetilde{\omega}^2\right)^2,&~& \qquad b^{II}_4= \frac{1}{280}\left(\widetilde{k}^2-\widetilde{\omega}^2\right)^2, \ldots \; .
\end{eqnarray}
Close to the horizon at $\widetilde{y}=1$, we can find an incoming-wave solution to eq.~(\ref{eq:Ezeqn}) of the form
\begin{equation}\label{eq:Ehornum}
E(\widetilde{y})=f(\tilde{y})^{-\frac{i\tilde{\omega}}{6}} R(\widetilde{y}),
\end{equation}
where $R(\widetilde{y})$ is regular at the horizon.\\
In order to find the retarded two-point correlation functions $G^{tt}_R(\omega,k)$, $G^{xx}_R(\omega,k)$, and $G^{tx}_R(\omega,k)$, we merely need to compute the polarization function $\Pi(\omega,k)$, which can be easily achieved using the method outlined in \cite{Son:2002sd,Herzog:2002pc} (for a more general, and very useful,  prescription, cf. \cite{Kaminski:2009dh}): Essentially, one needs to functionally differentiate the on-shell boundary action with respect to the sources. \\
In terms of $\mathcal{A}_t$ and $\mathcal{A}_x$, the boundary term takes the following form
\begin{equation}
S_{\mathrm{B}} = \lim_{u\rightarrow u_B}\mathcal{N} \! \int d^4 x \bigg( f(u) g(u) \mathcal{A}_x \partial_u \mathcal{A}_x -g(u) f(u) G(u) \mathcal{A}_t \partial_u \mathcal{A}_t\bigg),
\end{equation}
where $u_B$ is some radial cut-off. In terms of $E(k_{\mu},u) $ (and $E(k_{\mu},y)$, resp.), we obtain
\begin{align}\label{actionE}
S_{\mathrm{B}}& = \lim_{u\rightarrow u_B}\mathcal{N} \! \int \frac{d^4 k}{(2\pi)^4}\,\left( \frac{f(u) g(u) G(u)}{k^2-\omega^2 G(u)}\right)\, E(-k_{\mu}, u) \,\partial_u \,E(k_{\mu},u),\\
&=  \lim_{y\rightarrow 0} 2 \mathcal{N} R^{9/2}\! \int \frac{d^4 k}{(2\pi)^4}\,\left( \frac{h_2(y)}{k^2-\omega^2 G(y)}\right)\, E(-k_{\mu}, y) \,\partial_y \,E(k_{\mu},y).
\end{align}
Now, close to the boundary, we can expand
\begin{equation}
 \lim_{y\rightarrow 0} E(-k_{\mu}, y) \,\partial_y \,E(k_{\mu},y) = \mathcal{A}(-k_{\mu}) \mathcal{A}(k_{\mu})  \left( 2 b_2^I y + 3 \frac{\mathcal{B}}{\mathcal{A}} \frac{y^2}{y_H^3} + \ldots \right),
\end{equation}
where we have written the source of $E(\pm k_{\mu}, y \rightarrow 0)$ as $\mathcal{A}(\pm k_{\mu})$, in accordance with eq.~(\ref{eq:Ebdy}). As usual, one needs to holographically renormalize the action, i.e., remove the divergent terms from the action by adding appropriate counter terms:
\begin{align}
S_{\mathrm{B,ren.}} &= S_{\mathrm{B}} + S_{\mathrm{ct}}, \cr
S_{\mathrm{ct}} &=2 \mathcal{N} R^{9/2}\! \int \frac{d^4 k}{(2\pi)^4} \frac{\mathcal{A}(- k_{\mu})\mathcal{A}(k_{\mu})}{y}.
\end{align}
The matrix of (thermal) retarded correlators is given by
\begin{align}
G^{tt}_R(\omega,k) &= k^2 \,\Pi(\omega, k),\cr
G^{xx}_R(\omega,k) &= \omega^2 \,\Pi(\omega, k),\cr
G^{tx}_R(\omega,k) &= \omega\,k \,\Pi(\omega, k),\label{eq:Gxx}
\end{align}
with
\begin{align}\label{eq:defPi}
\Pi(\omega,k) \equiv \frac{\delta^2S_{\mathrm{B,ren.}}}{\delta E(-k, 0) \delta E(k,0)} \,.
\end{align}
Therefore, we arrive at
\begin{equation}\label{eq:Pi}
\Pi(\omega,k)= \frac{6 \mathcal{N} R^{9/2}}{y_H^3} \frac{1}{k^2- \omega^2} \frac{\mathcal{B}}{\mathcal{A}} =  \frac{16 \pi^3 \mu_8 \Omega_4}{9 g_s} R^6 T^3 \frac{1}{k^2- \omega^2} \frac{\mathcal{B}}{\mathcal{A}}.
\end{equation}
The coefficients $\mathcal{A}$ and $\mathcal{B}$ can be obtained by matching the numerical solution satisfying incoming-wave boundary conditions at the horizon (\ref{eq:Ehornum}) with the solution (\ref{eq:Ebdy}) valid near the boundary.\\
In conclusion, the longitudinal retarded Green's function and density-density correlation function are given by
\begin{align}
G_R^{xx}(\omega,k)&=  \frac{6 \mathcal{N} R^{9/2}}{y_H^3} \frac{\omega^2}{k^2- \omega^2} \frac{\mathcal{B}}{\mathcal{A}},\\
G_R^{tt}(\omega,k)&= \frac{6 \mathcal{N} R^{9/2}}{y_H^3} \frac{k^2}{k^2- \omega^2} \frac{\mathcal{B}}{\mathcal{A}} = \frac{k^2}{\omega^2}G_R^{xx}(\omega,k),
\end{align}
and we note that the two correlation functions share the same poles. Moreover, the spectral functions of the corresponding operators can be straightforwardly obtained as
\begin{equation}\label{eq:spectral}
\chi_{xx}(\omega,k)= - 2 \, \mathrm{Im}\left[G_R^{xx}(\omega,k)\right], \quad \chi_{tt}(\omega,k)= - 2\,  \mathrm{Im}\left[G_R^{tt}(\omega,k)\right].
\end{equation}
We will study these quantities numerically in the next section.
\section{Numerical results}

The main goal of the present work is to investigate the behavior of the collective modes at non-zero temperatures. Based on intuition from Landau-Fermi liquids and the $D3/D7$ model, for $T \ll \mu$, we expect to be able to roughly distinguish between three different regimes, with increasing temperature:
\begin{itemize}
\item region I (collisionless quantum): $\frac{T_H}{\hat{\mu}} \ll \left( \overline{\omega}, \overline{k}\right) \ll 1$, \\ or $1 \ll \left( \widetilde{\omega}, \widetilde{k} \right) \ll \widetilde{\mu} \sim d^{1/5}$,
\item region II (collisionless thermal): $ \left(\frac{T_H}{\hat{\mu}}\right)^{2} \ll \left(\overline{\omega}, \overline{k} \right)\ll  \left(\frac{T_H}{\hat{\mu}}\right)$, \\or $\widetilde{\mu}^{-1}\sim d^{-1/5} \ll \left( \widetilde{\omega}, \widetilde{k} \right) \ll 1 $,
\item region III (hydrodynamic): $ 0 \leq \left(\overline{\omega}, \overline{k}\right) \ll  \left(\frac{T_H}{\hat{\mu}}\right)^2$, \\or $ 0 \leq \left( \widetilde{\omega}, \widetilde{k}\right) \ll \widetilde{\mu}^{-1} \sim d^{-1/5}$.
\end{itemize}
This structure of regimes is expected only for weakly coupled, weakly interacting Fermi liquids, and it is clear that the behavior will be different in our holographic model of a quantum liquid at strong coupling. As we will see below, it can still serve as a useful guideline to classify the temperature-dependent changes that the system undergoes, even though the exact temperature dependence of the damping rate $\Gamma_T$ will be different in our model compared to the $D3/D7$ model and Landau-Fermi theory. It is also possible to numerically evolve the system deep into the high temperature regime, where the
zero sound mode has completely disappeared, and to study the diffusion mode in this regime. Our  results are in very good agreement with previous results for the diffusion mode in the Sakai-Sugimoto model (see \cite{Kim:2008bv}).\\ 
We are mostly interested in elucidating the temperature dependence of the relevant quantities and compare our numerical results with expectations from Landau-Fermi liquid theory and related holographic models based on $D3/D7$ brane constructions \cite{Bergman:2011rf, Davison:2011ek}. We verify that the zero sound mode continues to exist at finite (low) temperatures. Increasing the temperature eventually destabilizes the zero sound mode and the diffusive mode begins to dominate the density-density correlator.

\subsection{Quasi-normal modes and dispersion relations}
\begin{figure}[h] 
   \centering
     \includegraphics[width=4.0in]{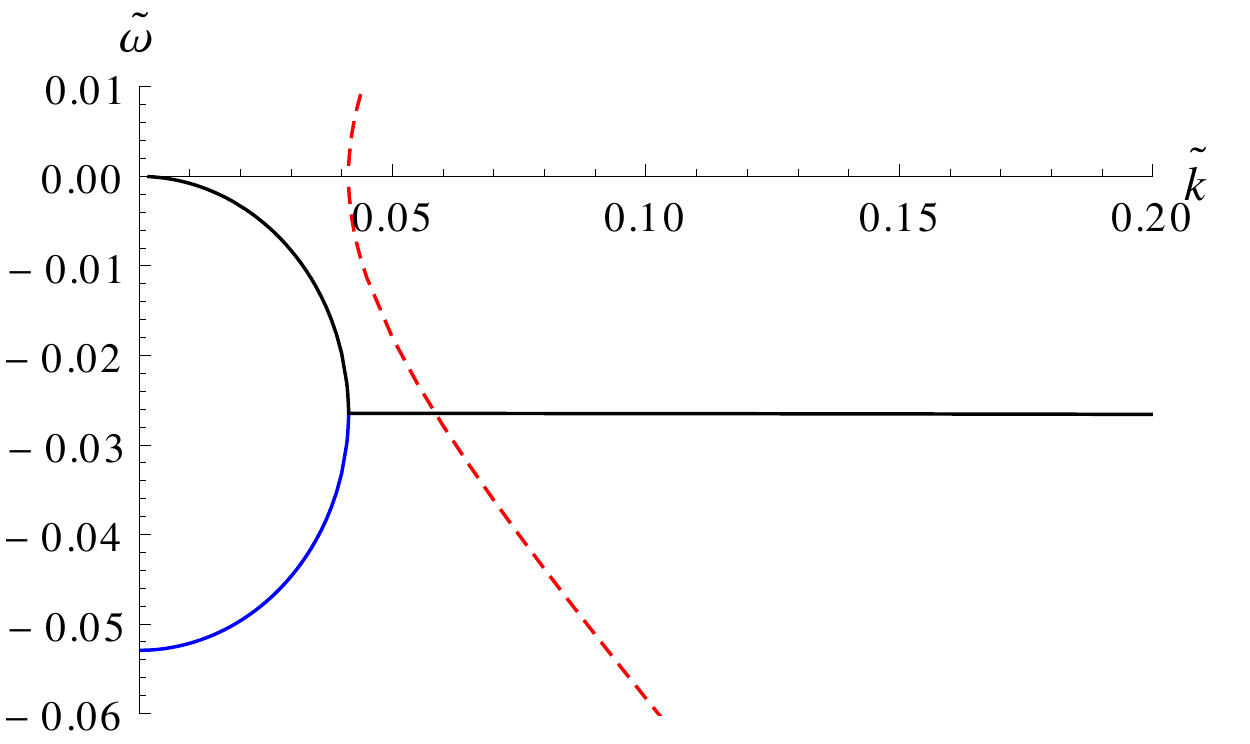}
    \caption{\small  Real (red dashed curve) and imaginary (blue and black solid curve) parts of the dispersion relation $\tilde{\omega}(\tilde{k})$ for $\tilde{\mu}=5$. The black solid curve indicates the more stable, longer-lived  solution, with less negative imaginary part.}
   \label{fig:qnmplot1}
\end{figure}

\begin{figure}[h] 
   \centering
      \includegraphics[width=4.0in]{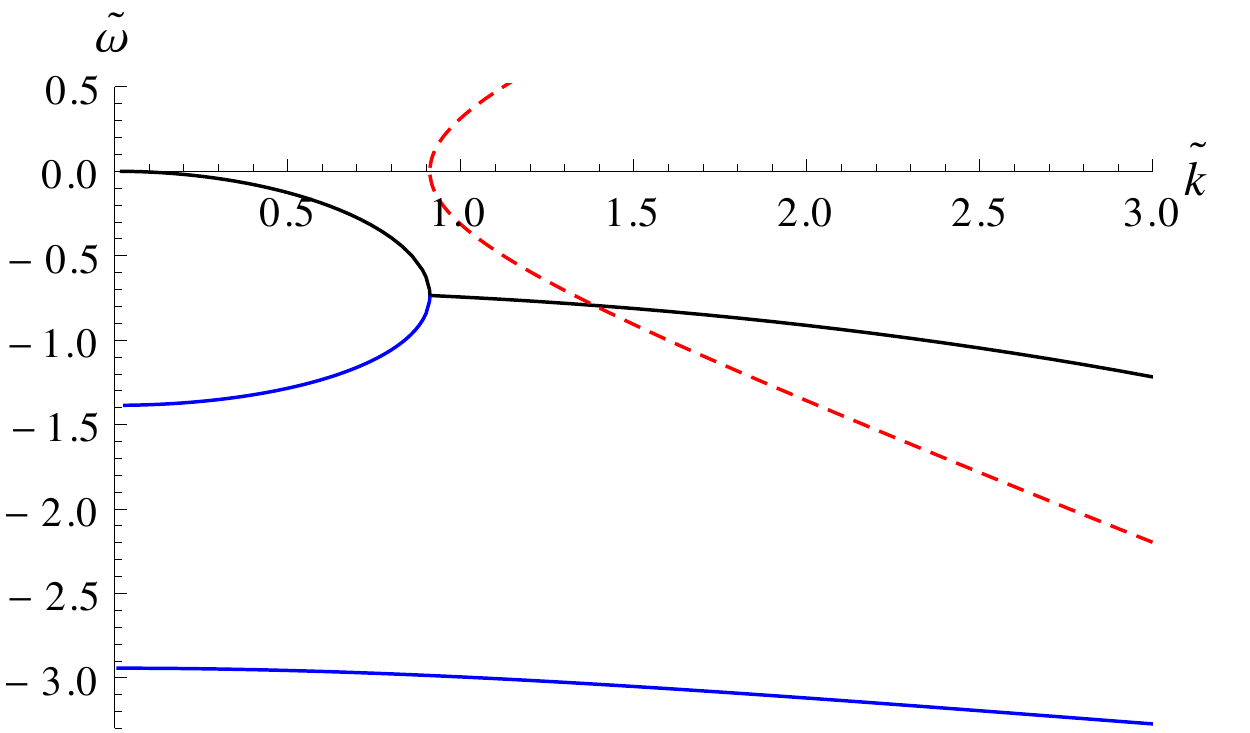}
    \caption{\small  Real (red dashed curve) and imaginary (blue and black solid curve) parts of the dispersion relation $\tilde{\omega}(\tilde{k})$ for $\tilde{\mu}=1.2$. The black solid curve indicates the more stable, longer-lived solution, with less negative imaginary part.}
   \label{fig:qnmplot2}
\end{figure}

\begin{figure}[h] 
   \centering
      \includegraphics[width=4.0in]{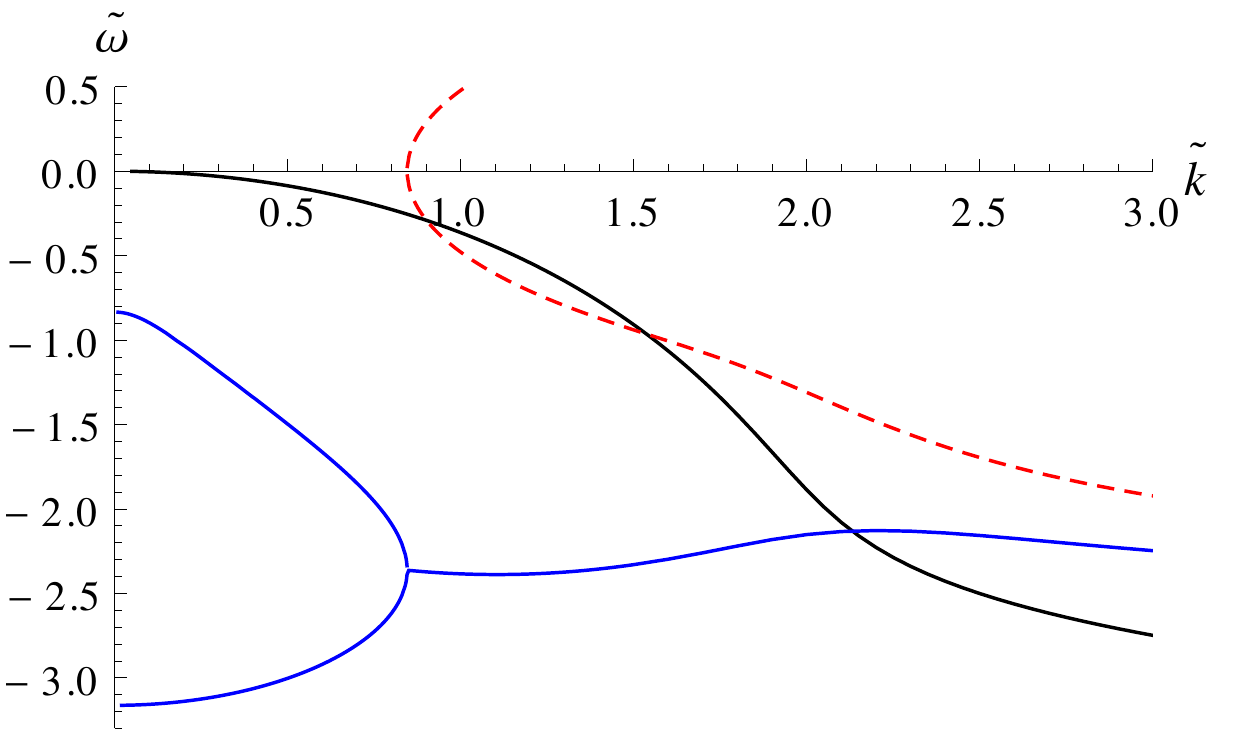}
    \caption{\small  Real (red dashed curve) and imaginary (blue and black solid curve) parts of the dispersion relation $\tilde{\omega}(\tilde{k})$ for $\tilde{\mu}=0.8$.  At large $\tilde{k}$, the zero sound mode (blue solid curve) is the most stable one, but for lower $\tilde{k}$ another quasi-normal mode (black solid curve) becomes more stable.}
   \label{fig:qnmplot3}
\end{figure}

The poles of the retarded Green's functions (\ref{eq:Gxx}) will be determined by the zeroes of $\mathcal{A}(k_{\mu})$, according to (\ref{eq:Pi}).
Since $\mathcal{A}(k_{\mu})$ is the source of $E(k_{\mu}, y)$ satisfying incoming-wave boundary conditions at the horizon, those $E(k_{\mu}, y)$ for which $\mathcal{A}(k_{\mu})$ vanishes precisely correspond to the quasi-normal modes of the system. From the dominant poles of the correlation functions (for fixed parameters $T$ and $\mu$, and a given $k$), one can thus obtain the dispersion relation $\omega (k)$, chosen such that the solution $E(k_{\mu}, y)$ vanishes at the boundary.\\
There are two limits, in which analytical results can be obtained:
In \cite{Kulaxizi:2008jx}, the dispersion relation at zero temperature was found to be
\begin{align}\label{eq:dispzero}
\omega (k)\Big{|}_{T_H=0} &= \omega_{\mathrm{R},0}(k)- \mathrm{i} \Gamma_0, \quad \mathrm{where} \cr
 \omega_{\mathrm{R},0}(k)&= \pm \sqrt{\frac{2}{5}}k +\frac{-1+2 \gamma + \ln \frac{k^2}{10}}{10^{3/2} \hat{\mu}^2}k^3 \quad \mathrm{and} \quad \Gamma_0=  \frac{\pi}{10^{3/2}\hat{\mu}^2}k^3 + \mathcal{O}\left(k^5\right),
\end{align}
where $\gamma \approx 0.577216$ is the Euler-Mascheroni constant, and $\Gamma_0$ is the attenuation (damping) rate at zero temperature.
This expression is peculiar due to the non-standard $k$-dependence of the imaginary part which is inconsistent with Landau-Fermi liquid theory.
Recall, however, that the $T=0$ result is thermodynamically irrelevant, since the formation of stripes \cite{Ooguri:2010xs} is expected to set in before then.
The other case in which one can get an analytic result is the high temperature limit 
(for large and small densities $\hat{\mu}$ resp., cf. \cite{Kim:2008bv}, eq. (101)):
\begin{equation}\label{eq:displarge}
\omega (k)\Big{|}_{T_H \rightarrow \infty} \sim \left\{ \begin{matrix} -i \frac{k^2 \hat{\mu}^2}{2 \pi T^3_H} \\ - i \frac{k^2}{2 \pi T_H} \end{matrix} \right. ,
\end{equation}
which is valid for small $\omega, k$. \\ 
In the following, we are going to scrutinize the non-zero temperature dependence of the dispersion relation and in particular exhibit the temperature dependence of its imaginary part. \\
To this end, we will utilize the numerical method applied in \cite{Bergman:2011rf} 
to the case at hand.  We choose three typical values for $\tilde{\mu}$ to illustrate our results on the fate of the zero sound mode with increasing temperature (see fig. \ref{fig:qnmplot1} -- \ref{fig:qnmplot3}). One can see that for every finite $\tilde{\mu}$, i.e., non-zero temperature, both a zero sound and a fundamental matter diffusion mode are present. Let us point out that there are no unstable modes (with positive imaginary part) for any value of $\tilde{\mu}$ in this sector.
\begin{figure}[h] 
   \centering
      \includegraphics[width=4.0in]{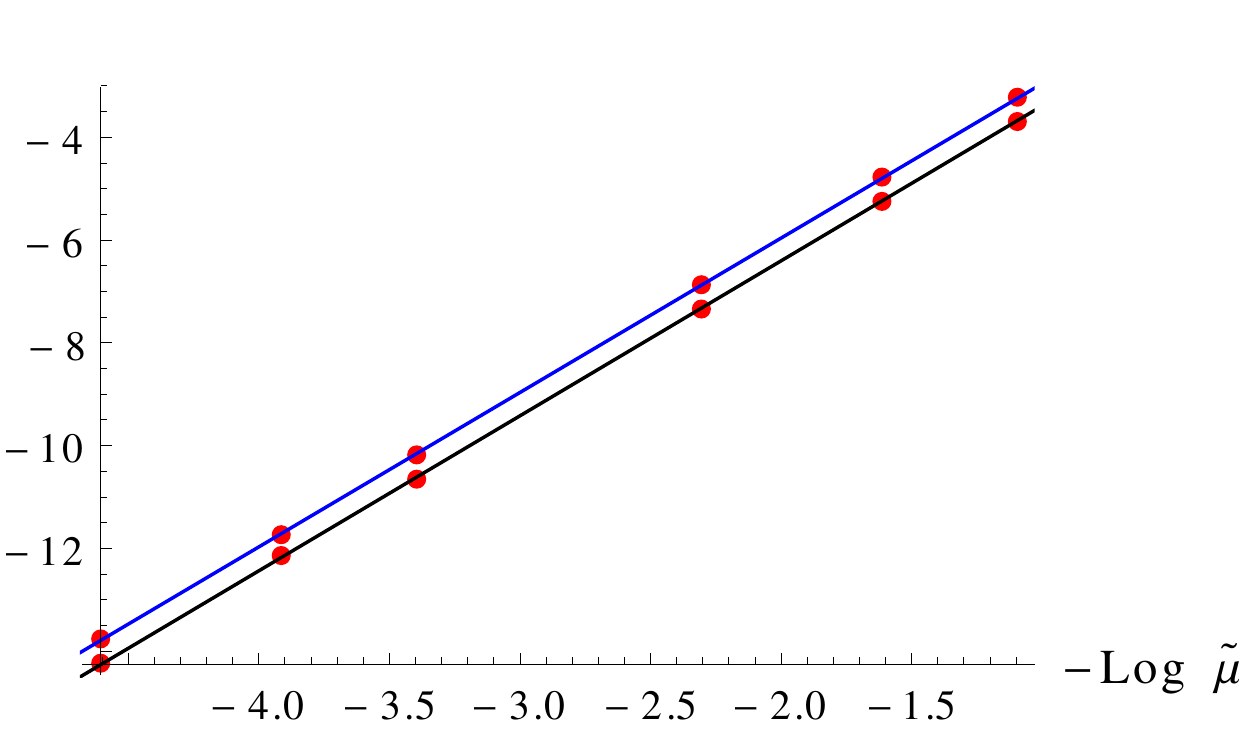}
    \caption{\small  The two lines correspond to $\log \bar k$ (upper blue) and $\log|\bar\omega|$ (lower black) at the crossover transition point from the collisionless to the hydrodynamical region. From the best fits to lines we infer that the temperature scaling is $\sim T^3$.}
   \label{fig:transitionpoint}
\end{figure}
For large $\tilde{k}$, there is a zero sound mode which continues to dominate for decreasing $\tilde{k}$ until, for some critical $\tilde{k}$ (depending on $\tilde{\mu}$), the real part of the dispersion relation vanishes and the
zero sound mode ceases to propagate. Below this threshold, there are two branches of purely imaginary modes, one of which is more stable and longer-lived than the other (the one closer to the real axis). 
This is the diffusion mode that vanishes at $\tilde{k}=0$. The location where the two purely imaginary modes merge and form a pair of complex modes (which we have identified as the non-zero temperature zero sound modes),
will be referred to as the crossover transition point.
We find that the crossover location scales as $T^3$, i.e., the values of $\bar\omega_{crossover},\bar k_{crossover}\propto T^3$; this is depicted in fig.~\ref{fig:transitionpoint}.
Put differently, starting at very low temperatures, cf. fig. \ref{fig:qnmplot1},
and decreasing $\tilde{\mu}$, corresponding to raising the temperature for fixed baryon density, the critical $\tilde{k}$ increases, meaning that the zero sound mode stops propagating at larger $\tilde{k}$ (cf. fig. \ref{fig:qnmplot2}).
Another, less stable, quasi-normal mode appears. For even smaller $\tilde{\mu}$ (fig. \ref{fig:qnmplot3}), this quasi-normal mode becomes more stable than the zero sound and the purely imaginary modes below a certain value of $\tilde{k}$.

\subsubsection{Behavior of attenuation rate $\Gamma_T$ at non-zero temperature}

\begin{figure}[h] 
   \centering
      \includegraphics[width=5.0in]{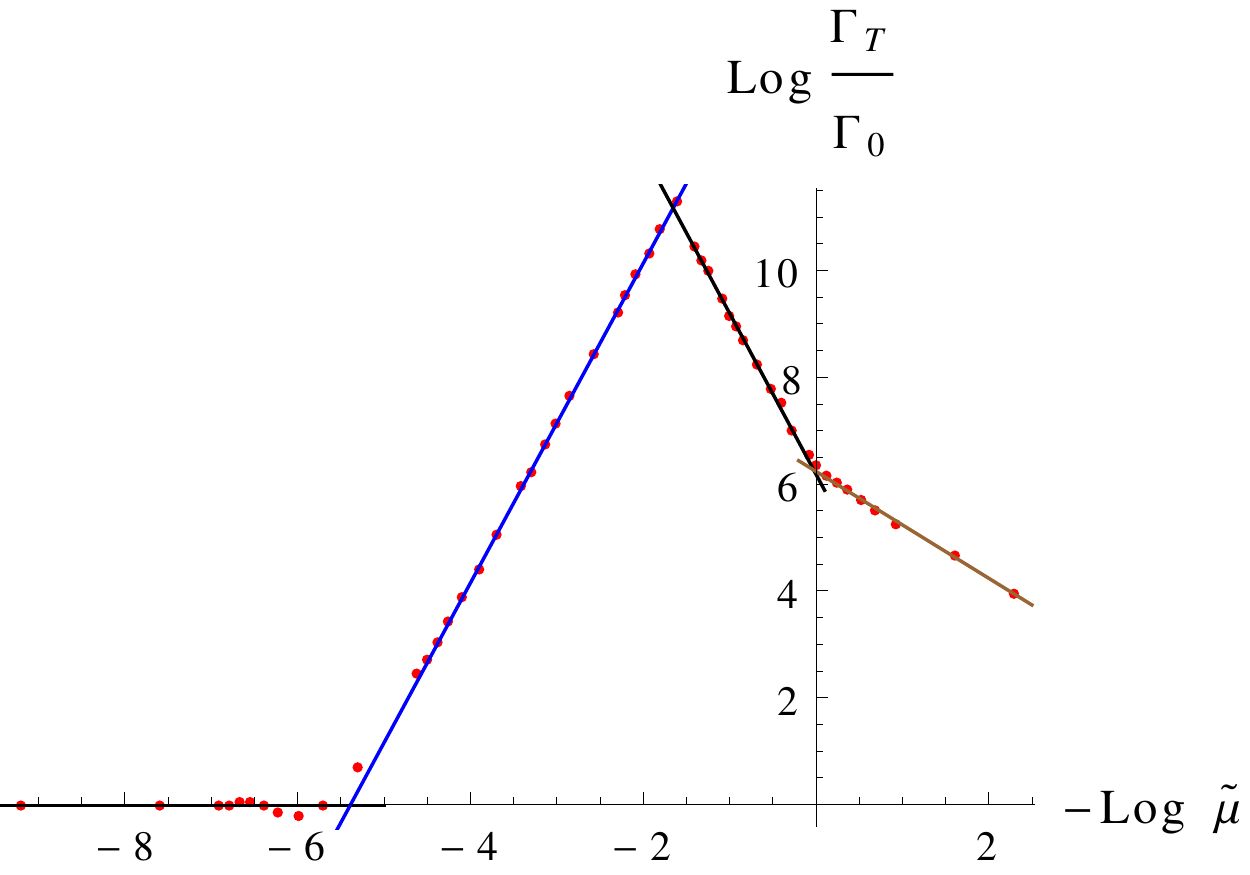}
  \caption{\small Log-log plot of $\Gamma_T$ (normalized by $\Gamma_0$) versus $\tilde{\mu}^{-1}$ for  $\bar k=0.01$. There are four different regimes. The two leftmost regions correspond to zero sound and the two rightmost regions correspond to a diffusive mode. From the respective slopes of the line segments, one can infer the temperature dependences of the different regions.}
   \label{fig:Gammaplot}
\end{figure}

In the following we will numerically investigate the behavior of the attenuation or decay rate. We will be able to distinguish between different regions and we will focus on the respective longest lived mode in each region.  Our results are presented in figure \ref{fig:Gammaplot}; they complement and extend the results obtained above for the dispersion relation.
Varying $\tilde{\mu}$, we can clearly distinguish between four separate regimes in which the behavior of the attenuation rate $\Gamma_T$ shows different temperature dependences.
In the plot, we are decreasing $\tilde{\mu}$ from left to right (increasing the temperature while keeping $\hat{\mu}$ fixed):
\begin{itemize}
\item region I (collisionless quantum): $\Gamma_T$ is a constant (independent of $T_H$) and essentially the same as for zero temperature, $\Gamma_0$, cf. eq. (\ref{eq:dispzero}).
\item region II (collisionless thermal): $\Gamma_T \sim T_H^3$.
\item region IIIa (cold hydrodynamical): $\Gamma_T \sim T_H^{-3}$, in complete agreement with the upper eq. in (\ref{eq:displarge}).
\item region IIIb (hot hydrodynamical): $\Gamma_T \sim T_H^{-1}$, in complete agreement with the lower eq. in (\ref{eq:displarge}).
\end{itemize}
In regions I and II, the dominant mode is the complex zero sound mode, while in regions IIIa and IIIb, the dominant mode is a purely imaginary diffusive mode.

\subsection{Spectral functions}
\begin{figure}[h] 
   \centering
   \includegraphics[height=2.75in]{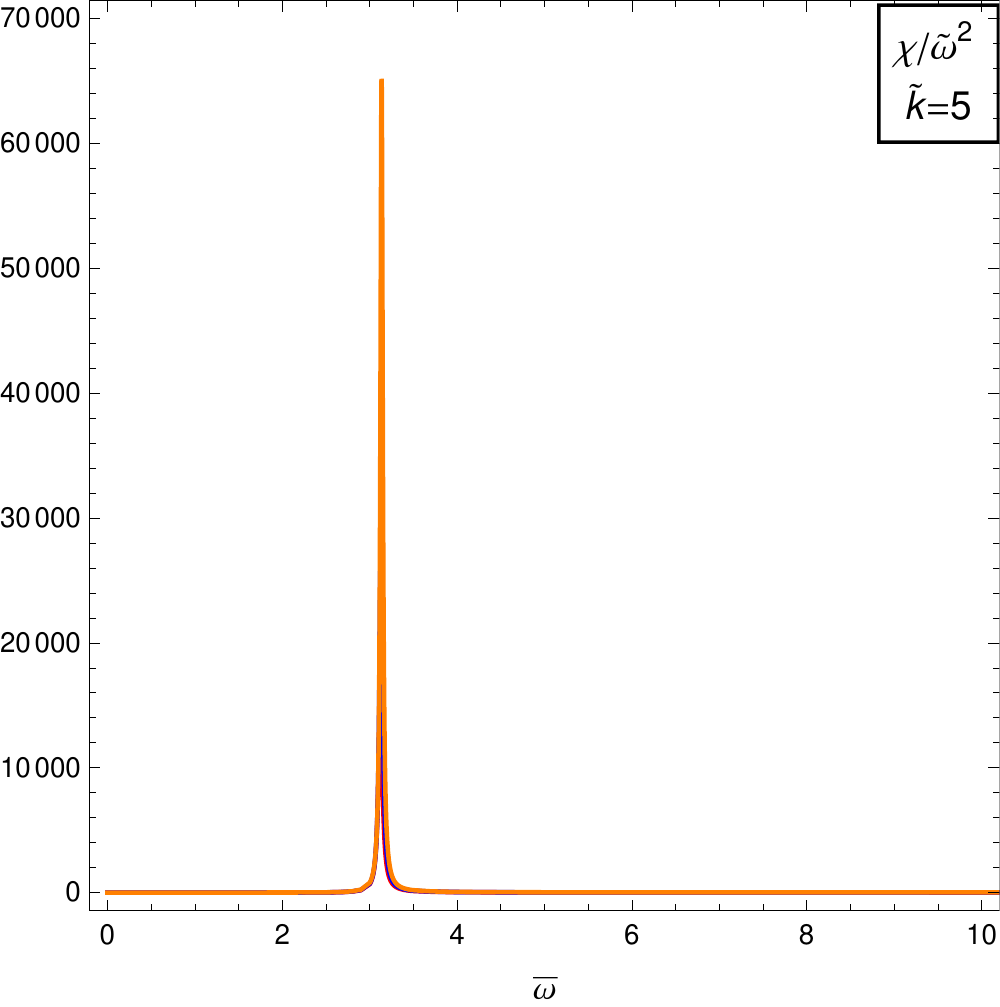}
 \includegraphics[height=2.75in]{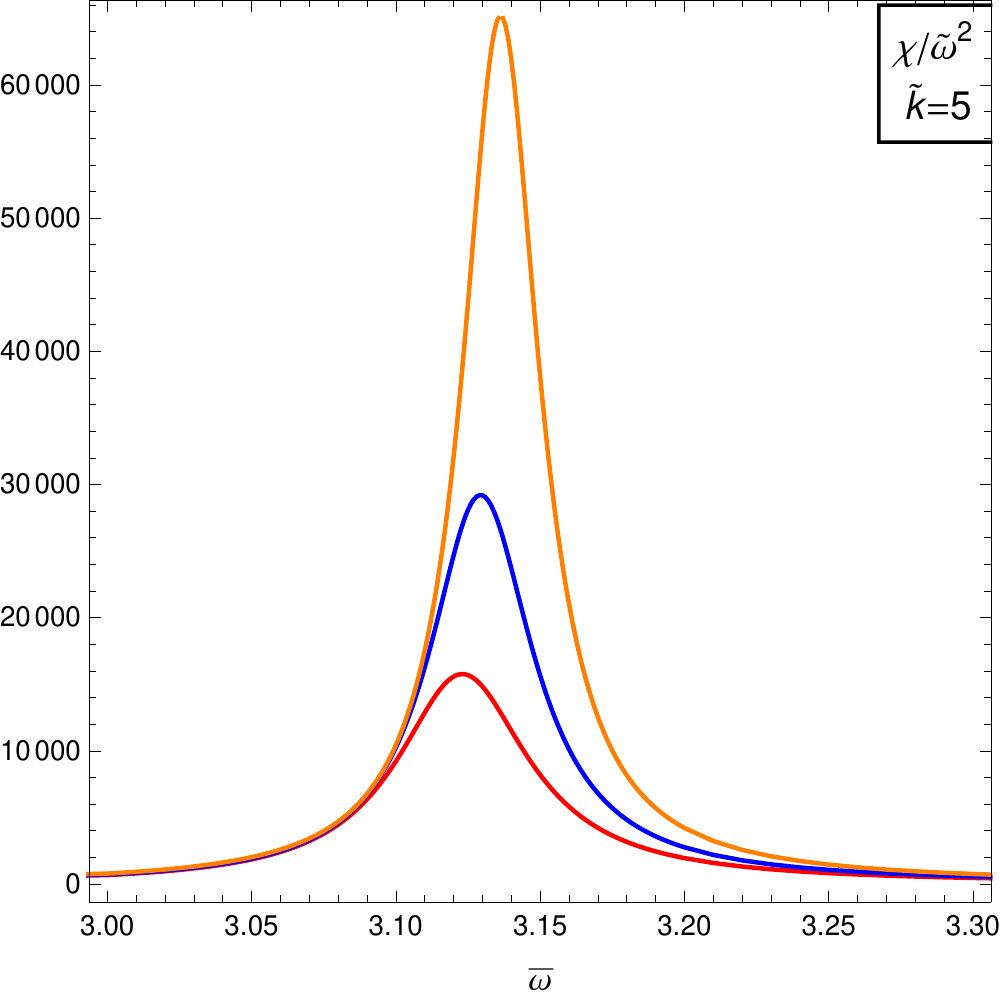}
    \caption{\small  Spectral function $\chi_{xx}(\tilde{\omega})$ (normalized by $\left( \frac{6 \mathcal{N} R^{9/2}}{y_H^3}\right) \tilde{\omega}^2$) in region I, collisionless quantum regime, for $\widetilde{k}=5$ and $\tilde{\mu}=10$ (red), $\tilde{\mu}=15$ (blue), $\tilde{\mu}=20$ (orange). Left panel: The sharp, distinct peaks correspond to the zero sound mode. Note that decreasing $\tilde{\mu}$ corresponds to raising the temperature, so that the highest peak occurs for the lowest temperature and largest $\tilde{\mu}$. There is no further significant structure in the spectral function at larger $\tilde{\omega}$. Right panel: Zoomed in closer to the proximity of the peaks. With decreasing $\tilde{\mu}$ (higher temperature), the location of the zero sound peak moves towards smaller $\tilde{\omega}$.}
   \label{fig:spectral1}
\end{figure}

\begin{figure}[h] 
   \centering
    \includegraphics[height=2.75in]{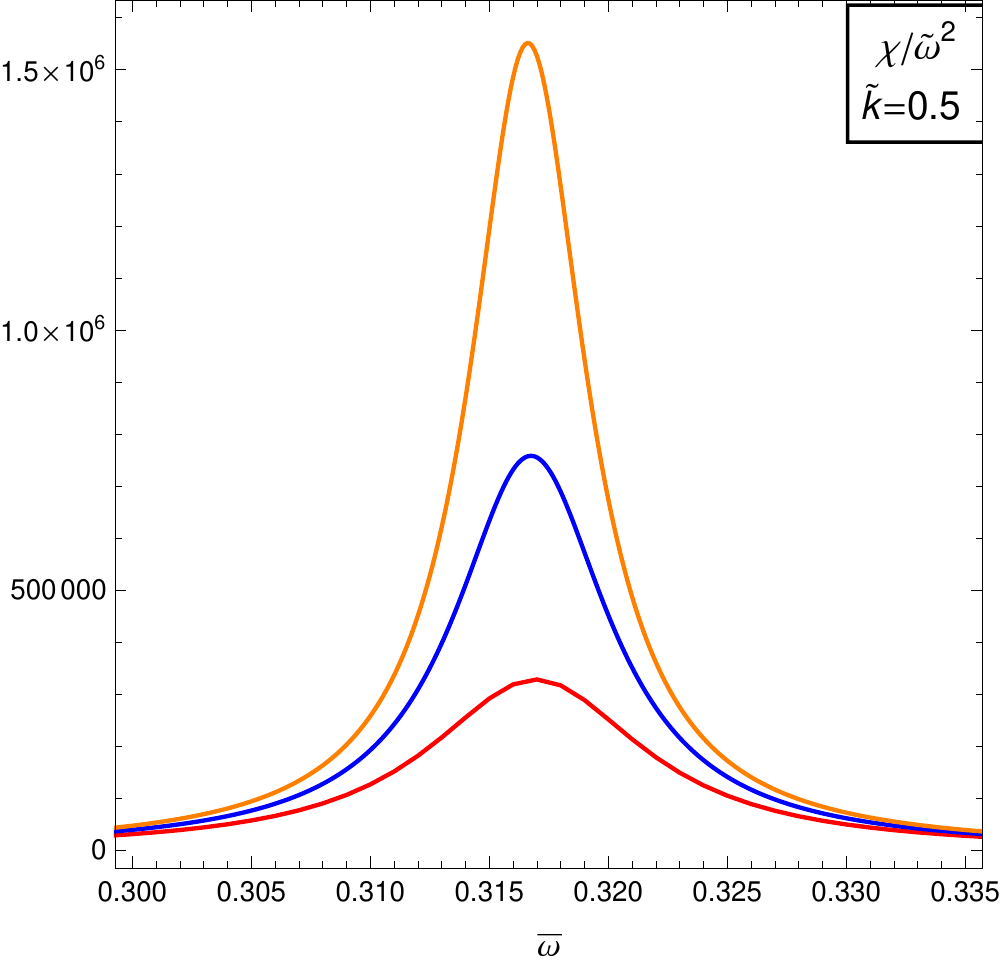}
     \includegraphics[height=2.75in]{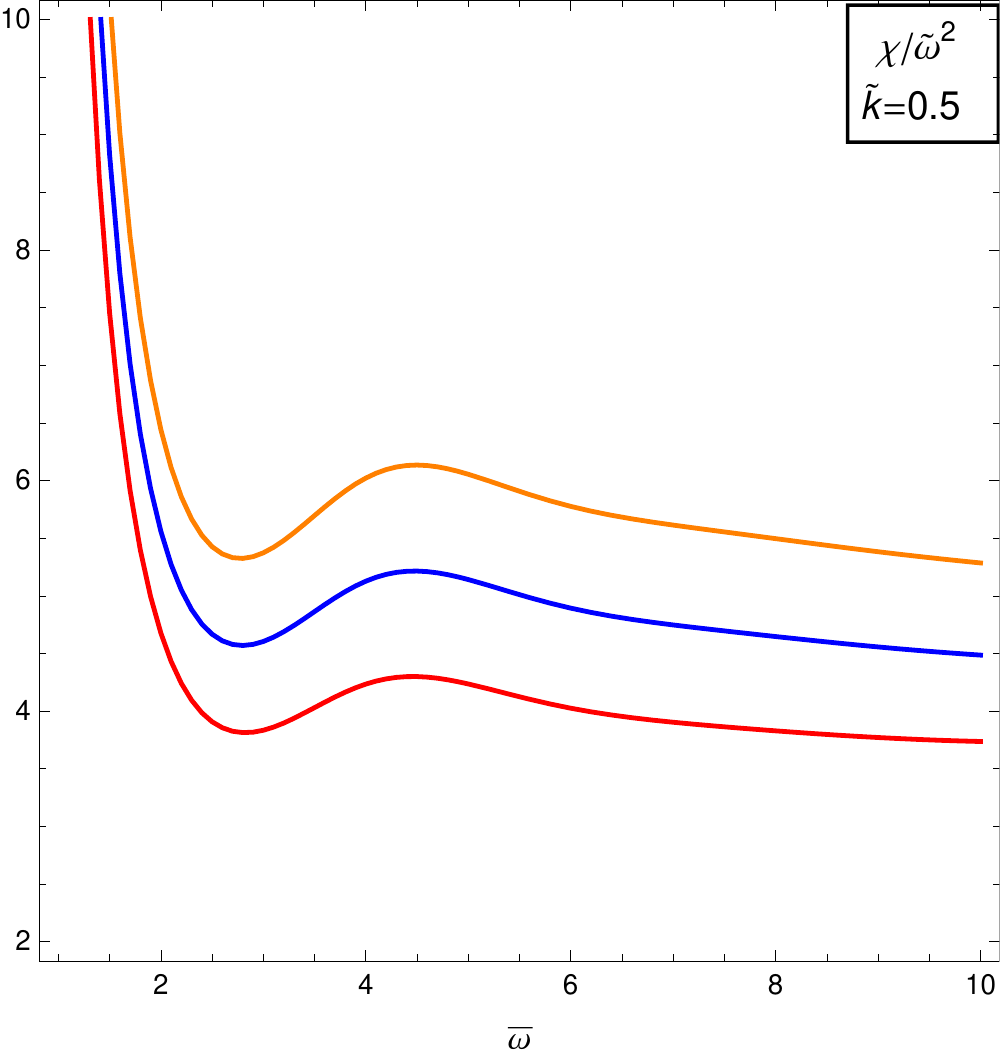}
    \caption{\small  Spectral function $\chi_{xx}(\tilde{\omega})$ (normalized by $\left( \frac{6 \mathcal{N} R^{9/2}}{y_H^3}\right) \tilde{\omega}^2$) in region II, collisionless thermal regime, for $\widetilde{k}=0.5$ and $\tilde{\mu}=11$ (red), $\tilde{\mu}=13$ (blue), $\tilde{\mu}=15$ (orange). Note that the normalization factor enhances the height of the peak for $\tilde{\omega}<1$. Left panel: Zoomed in closer to the proximity of the peaks. Right panel: At slightly higher $\tilde{\omega}$, there is a second peak developing as the temperature increases and thermal excitations become important.}
   \label{fig:spectral2}
\end{figure}

\begin{figure}[h] 
   \centering
      \includegraphics[height=2.75in]{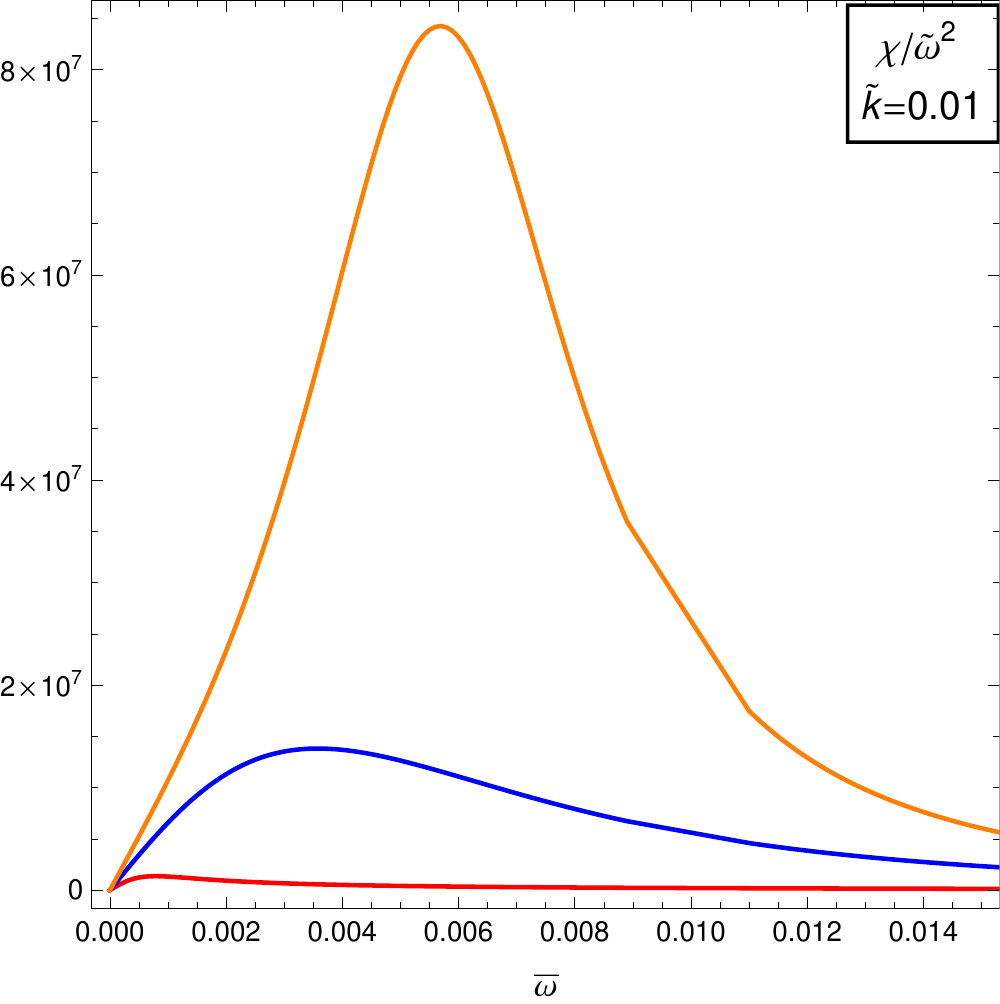}
      \includegraphics[height=2.75in]{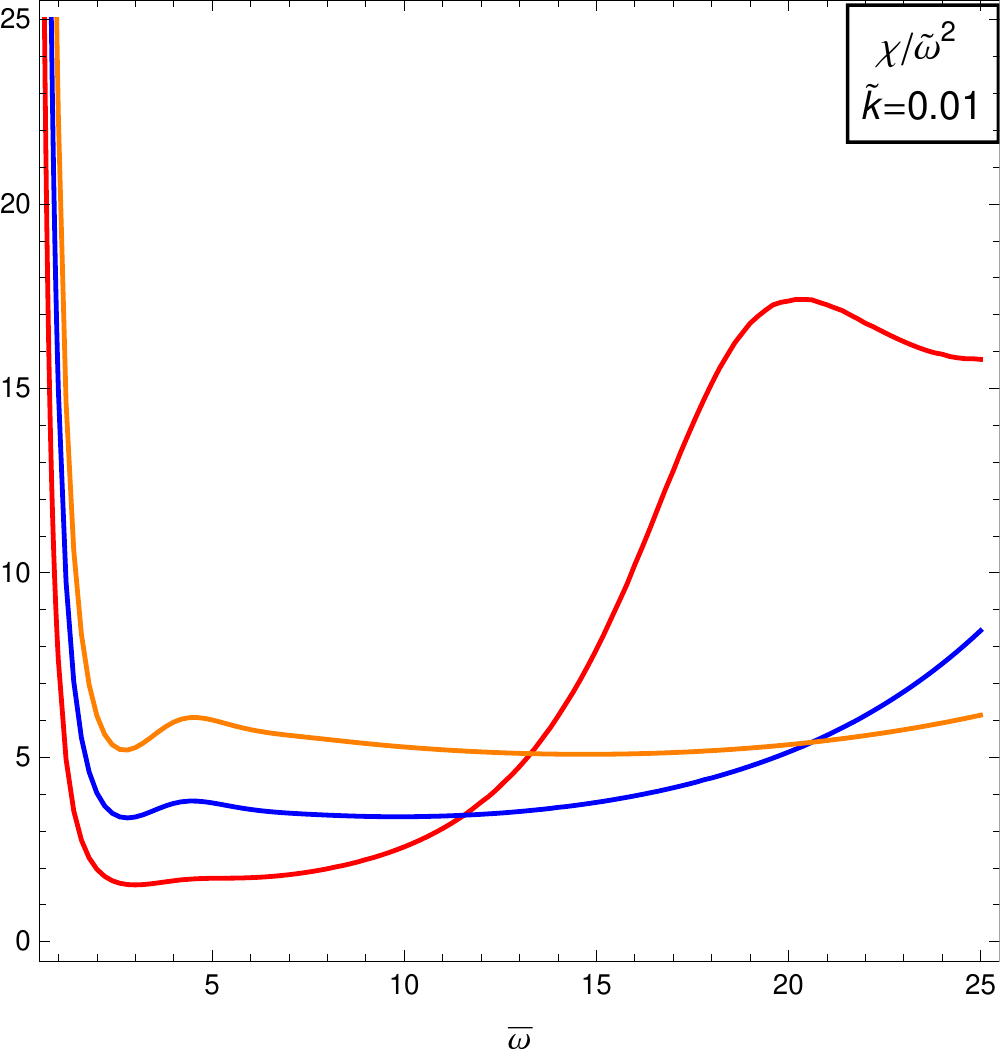}
    \caption{\small  Spectral function $\chi_{xx}(\tilde{\omega})$  (normalized by $\left( \frac{6 \mathcal{N} R^{9/2}}{y_H^3}\right) \tilde{\omega}^2$) in region III, hydrodynamic regime,  for $\widetilde{k}=0.01$ and $\tilde{\mu}=5$ (red), $\tilde{\mu}=10$ (blue), $\tilde{\mu}=15$ (orange). Left panel: The zero sound mode decays further and moves closer to the origin. Right panel: The hydrodynamic diffusive contribution becomes dominant compared to the remnants of the zero sound mode.}
   \label{fig:spectral3}
\end{figure}

\begin{figure}[h] 
   \centering
   \includegraphics[height=2.75in]{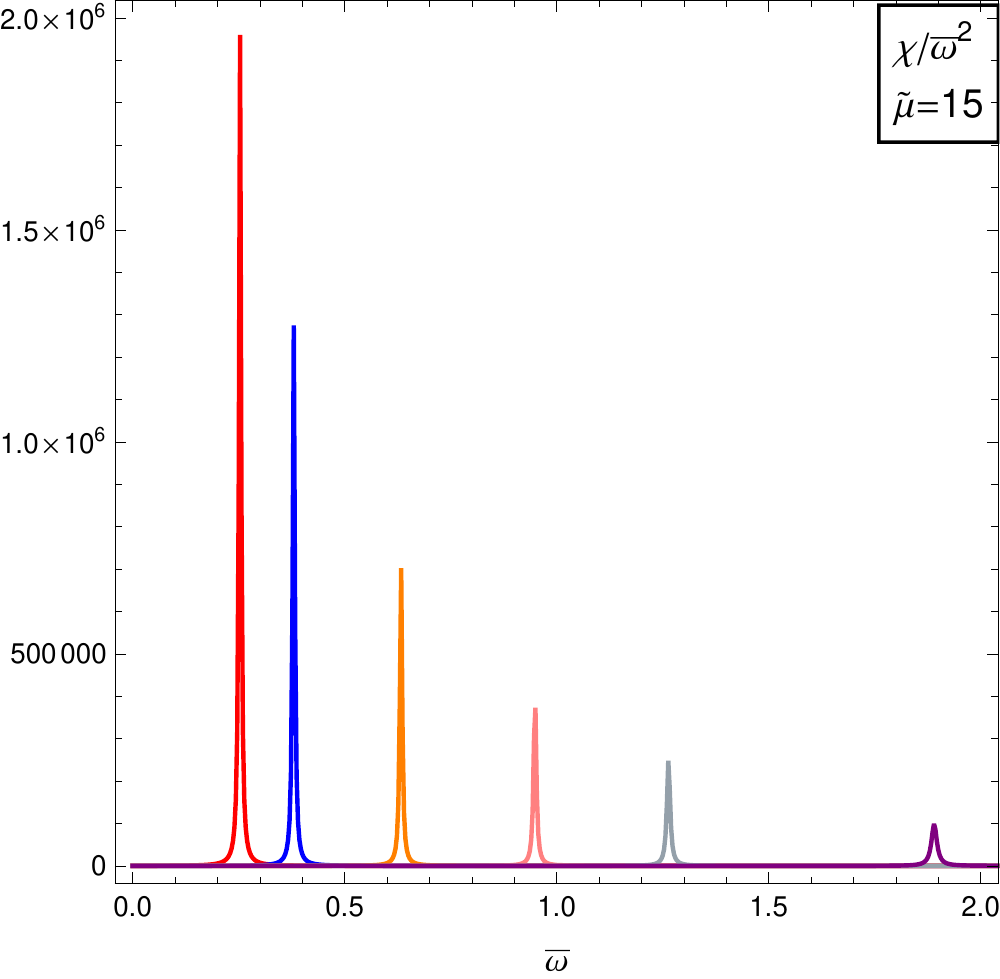}
   \includegraphics[height=2.75in]{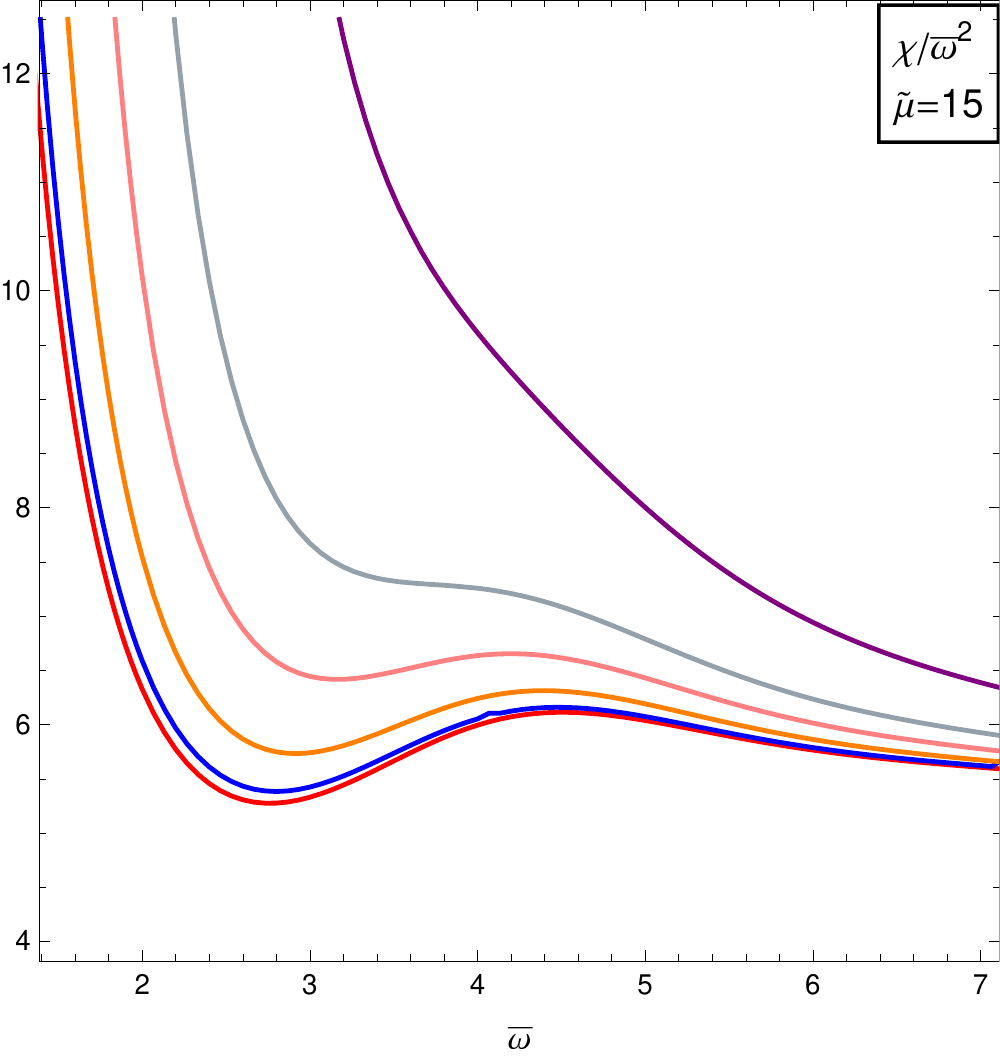}
    \caption{\small  Spectral function $\chi_{xx}(\overline{\omega})$ (normalized by $\left( \frac{6 \mathcal{N} R^{9/2}}{y_H^3}\right) \overline{\omega}^2$) in region I, collisionless quantum regime, for $\widetilde{\mu}=15$ and $\overline{k}=0.4$ (red), $\overline{k}=0.6$ (blue), $\overline{k}=1.0$ (orange), $\overline{k}=1.5$ (pink), $\overline{k}=2.0$ (grey), and  $\overline{k}=3.0$ (purple). Left panel: The sharp, distinct peaks correspond to the zero sound mode. Right panel: Zoomed in closer to the proximity of the peaks.}
 \label{fig:spectral1A}
\end{figure}

\begin{figure}[h] 
   \centering
   \includegraphics[height=2.75in]{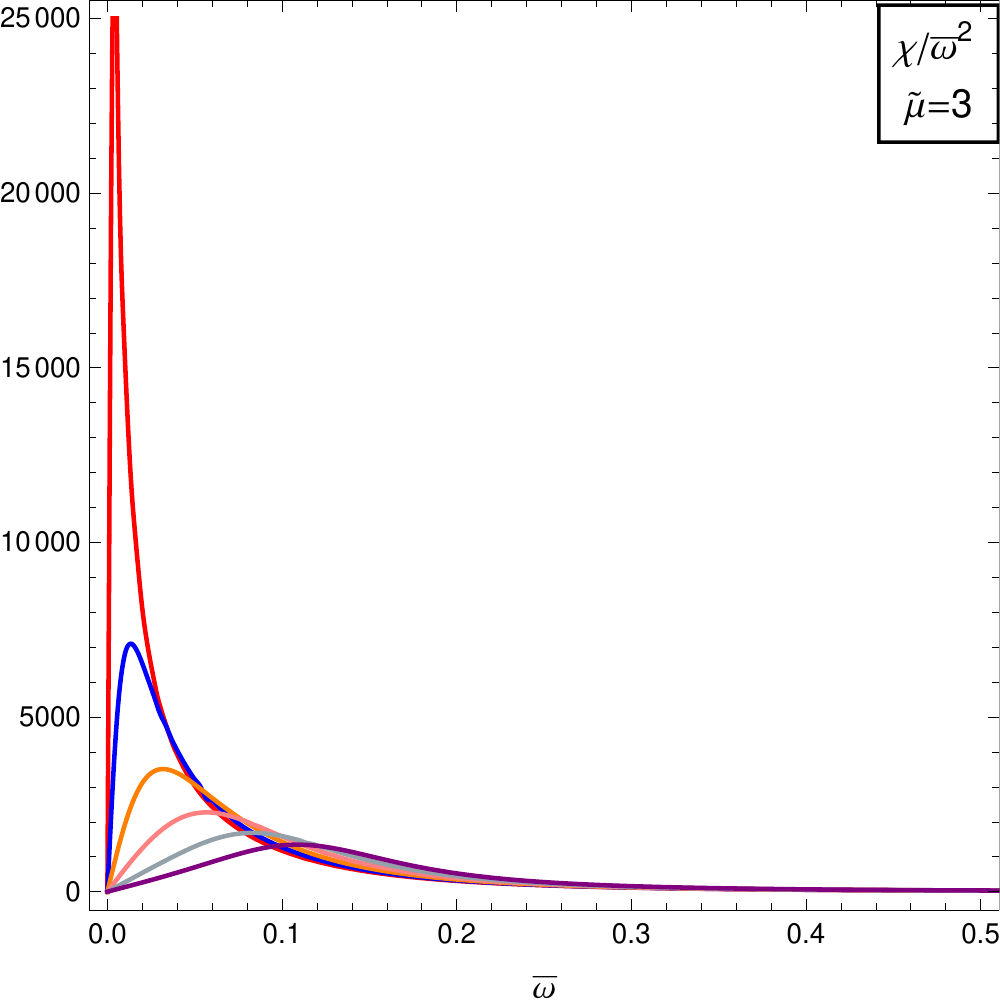}
 \includegraphics[height=2.75in]{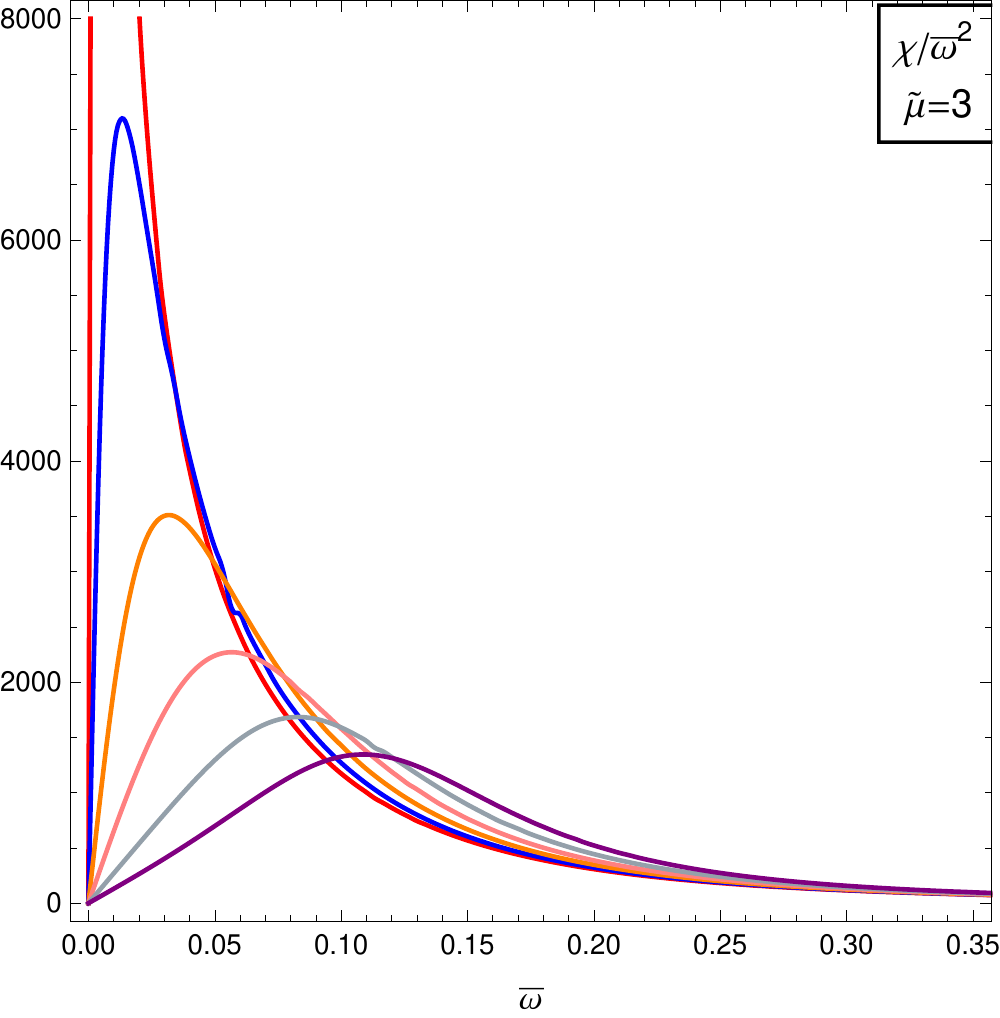}
    \caption{\small  Spectral function $\chi_{xx}(\overline{\omega})$ (normalized by $\left( \frac{6 \mathcal{N} R^{9/2}}{y_H^3}\right) \overline{\omega}^2$) in region II, collisionless thermal regime, for $\widetilde{\mu}=3$ and $\overline{k}=1/30$ (red), $\overline{k}=1/15$ (blue), $\overline{k}=1/10$ (orange), $\overline{k}=2/15$ (pink), $\overline{k}=1/6$ (grey), and  $\overline{k}=1/5$ (purple). Left panel: The zero sound mode decays and broadens. Right panel: Zoomed in closer to the proximity of the peaks.}
   \label{fig:spectral2A}
\end{figure}

\begin{figure}[h] 
   \centering
   \includegraphics[height=2.75in]{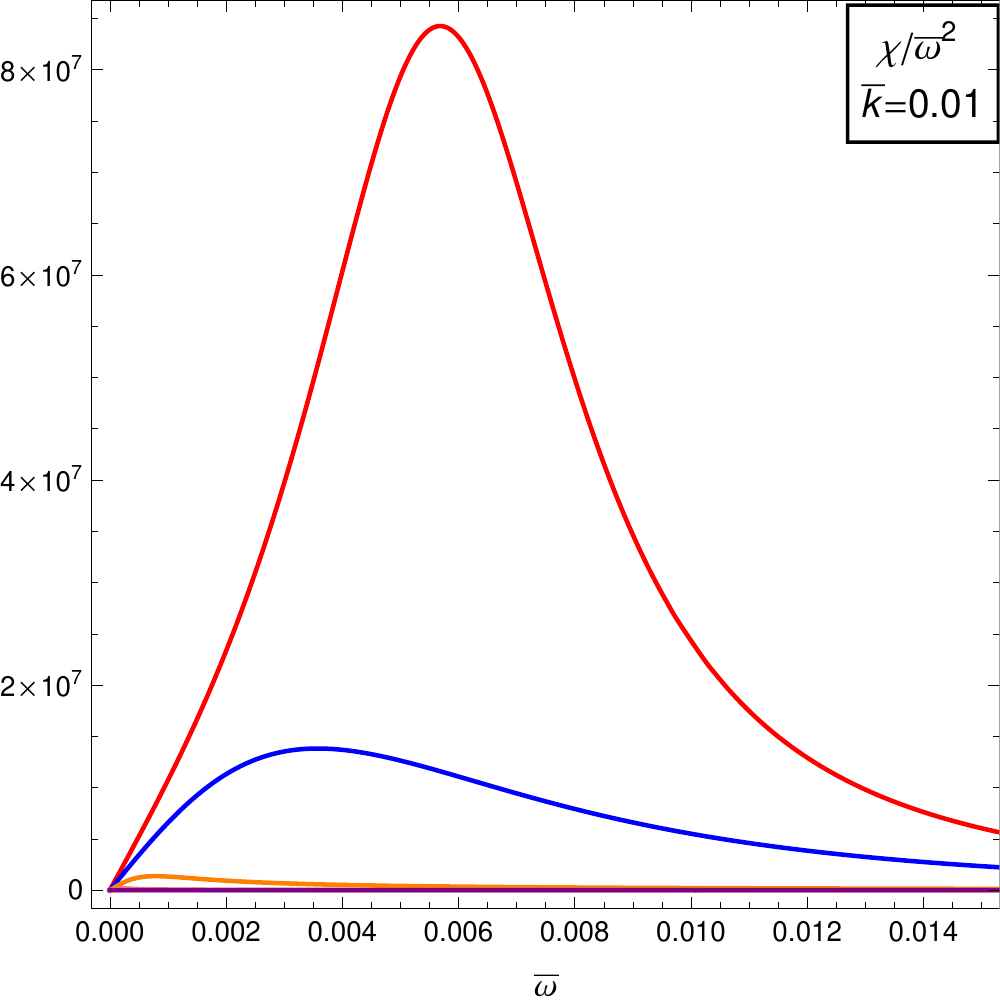}
 \includegraphics[height=2.75in]{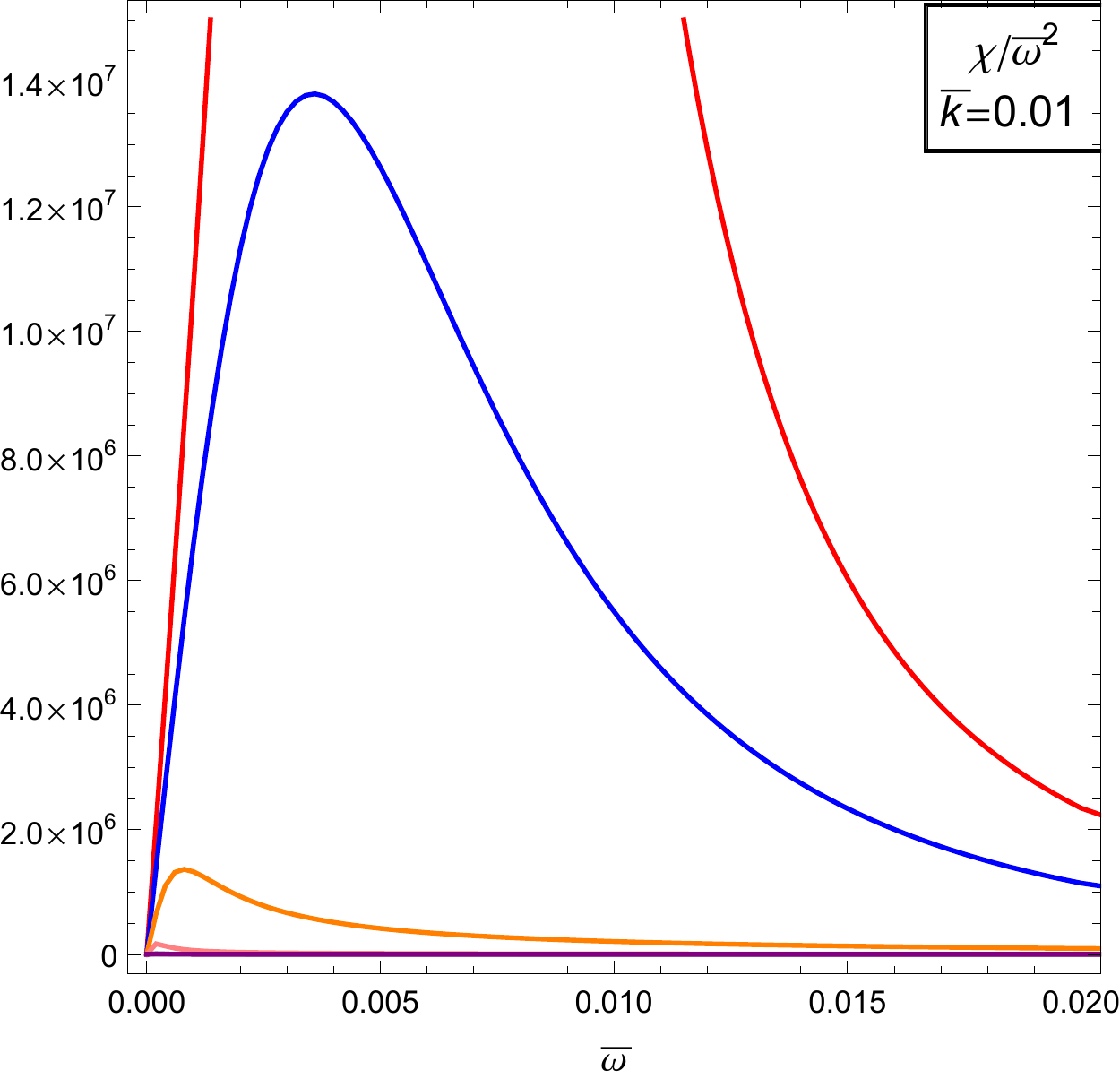}
 \includegraphics[height=2.75in]{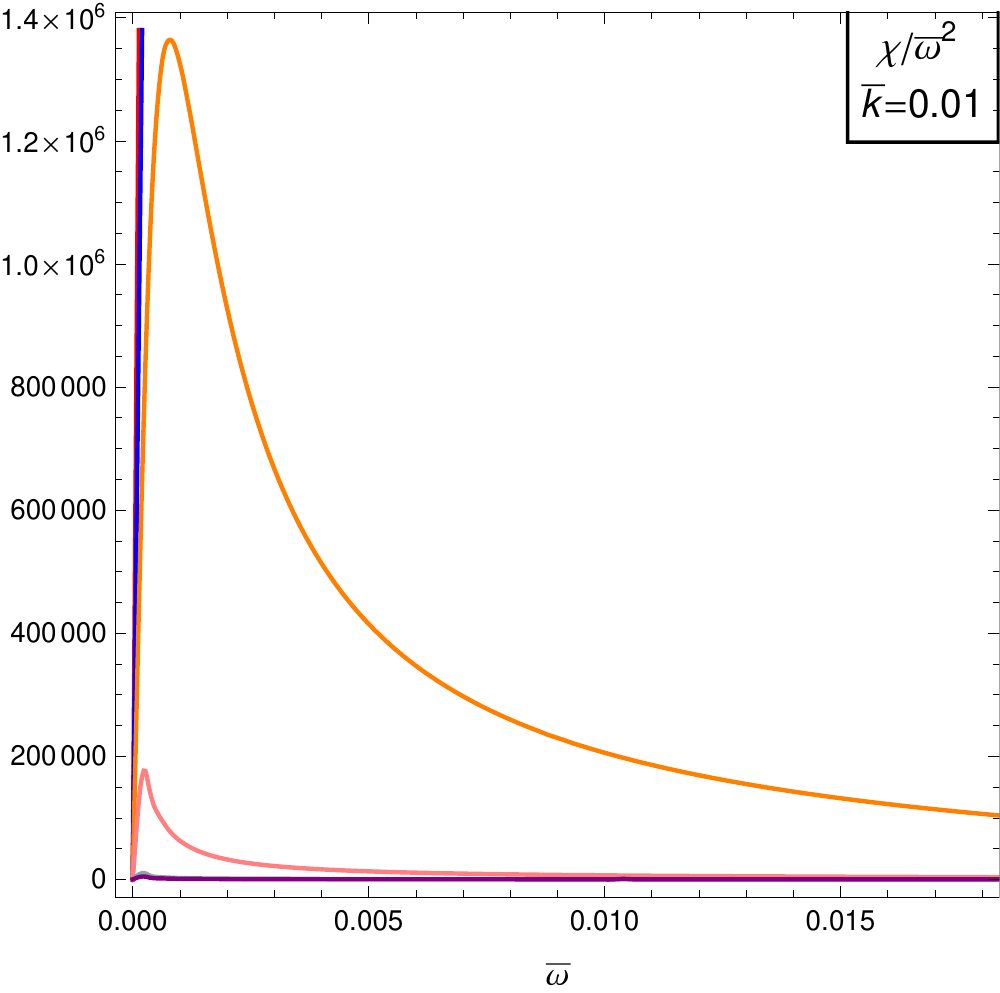}
    \caption{\small  Spectral function $\chi_{xx}(\overline{\omega})$ (normalized by $\left( \frac{6 \mathcal{N} R^{9/2}}{y_H^3}\right) \overline{\omega}^2$) in region III, hydrodynamic regime, for $\overline{k}=0.01$ and $\widetilde{\mu}=15$ (red), $\widetilde{\mu}=10$ (blue), $\widetilde{\mu}=5$ (orange), $\widetilde{\mu}=2.5$ (pink), $\widetilde{\mu}=1.25$ (grey), $\widetilde{\mu}=1$ (purple). Going from the largest to the smallest peak corresponds to increasing the temperature. Note the different scales on the vertical axis. Each subsequent figure zooms into an area of the previous figure (cf. \cite{Davison:2011ek}, fig. 10). The zero sound mode destabilizes as we move deeper into the hydrodynamic regime.}
   \label{fig:spectral3A}
\end{figure}

The spectral functions can be obtained via (\ref{eq:spectral}) from the retarded correlators. Here we will investigate the longitudinal component $\chi_{xx}(\tilde{\omega},\tilde{k})$, specifically the structure of the dominant peak
corresponding to the zero sound mode, its decay and the emergent diffusion peak as we increase temperature in going from region I to region III. Our numerical results are summarized in figures \ref{fig:spectral1}, \ref{fig:spectral2}, \ref{fig:spectral3},
where we plotted various spectral functions for values of $\tilde{k}$ and $\tilde{\mu}$ chosen to lie in the intervals defining regions I, II, and III (see definition above). The numerics get more involved and expensive for larger values of $\tilde{\mu}$, so we had to choose intermediate values that rendered the numerical computations feasible. Specifically, it turned out to be numerically advantageous to fix a value of $\tilde{k}$ and vary $\tilde{\mu}$ within the regional boundaries defined above.
Alternatively, we can employ the other set of dimensionless variables defined in eq. (\ref{eq:barvar}) to be able to compare more directly to plots in the existing literature, e.g., \cite{Davison:2011ek}. The relevant plots are presented in figure \ref{fig:spectral1A} (collisionless quantum regime), figure \ref{fig:spectral2A} (collisionless thermal regime), and figure \ref{fig:spectral3A} (hydrodynamic regime).

Qualitatively, the physical picture is quite similar to the $D3/D7$ model studied in \cite{Davison:2011ek}: The zero sound mode continues to dominate the spectral functions for low temperature (indicated by a sharp peak) in the collisonless quantum regime, i.e., region I. The height of the peak decreases with increasing temperature. In region II, the collisionless thermal regime, the decay of the zero sound mode is visible and progresses with a further increase in temperature, manifesting itself by a broadening peak that becomes less and less prominent. A second bump begins to develop at larger $\tilde{\omega}$.
Finally, when increasing temperature even higher into region III, the hydrodynamic regime, the zero sound mode decays further and moves closer to the origin. The second peak at larger $\tilde{\omega}$ that already started to appear in region II, becomes more and more dominant. This feature corresponds to thermal excitations and collisions, indicating the emergence of a hydrodynamic diffusion mode. Note that we are not able to visualise the hydrodynamic (first) sound mode in this framework, since it is an excitation of the energy momentum Green's function which is suppressed in the probe limit $N_f \ll N_c$.

\subsection{Absence of a Fermi surface}
\begin{figure}[h] 
   \centering
   \includegraphics[width=5.0in]{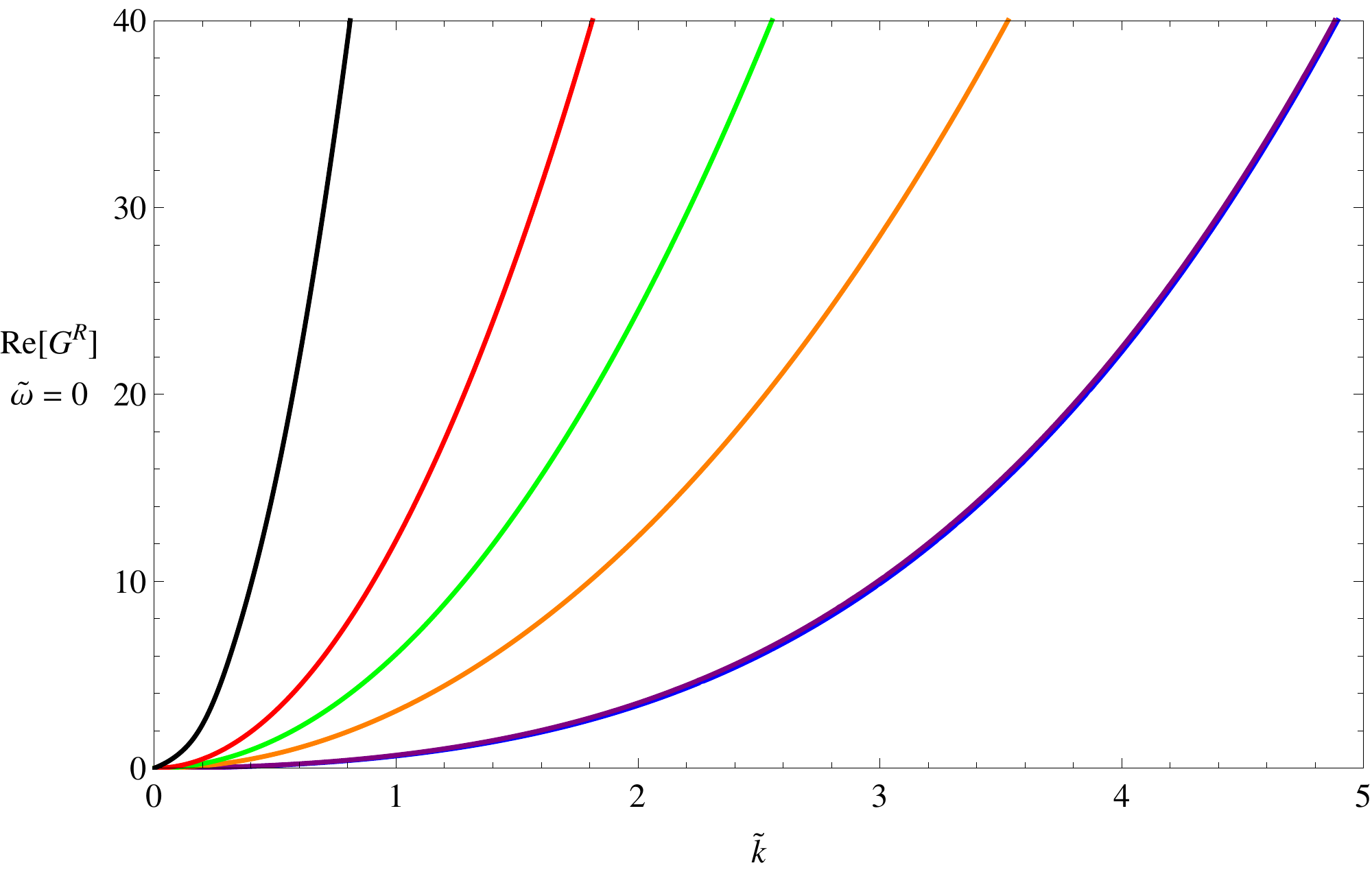}
    \caption{\small  The real part of $G_R(\widetilde{\omega}=0, \widetilde{k})$ for various choices of $\widetilde{\mu}$, from bottom to top, $\widetilde{\mu}=0$ (blue), $\widetilde{\mu}=1$ (purple), $\widetilde{\mu}=5$ (orange), $\widetilde{\mu}=10$ (green), $\widetilde{\mu}=20$ (red), $\widetilde{\mu}=100$ (black). Note that the $G_R(\widetilde{\omega}=0, \widetilde{k}=0)$ intercepts have been shifted to improve presentability.}
   \label{fig:ratio}
\end{figure}

This question was originally posed in ref.~\cite{Kulaxizi:2008jx} in the context of the zero temperature limit of the Sakai-Sugimoto model. Here we will review their arguments and extend the investigation to non-zero temperature.
From the theory of Fermi liquids, we know that the existence of a sharp Fermi surface is associated with a discontinuity in the distribution function which in turn is directly observable as a singularity at $k=2 k_F$ in the retarded current-current Green's functions in the small frequency limit,
\begin{equation}
G_R(\omega \rightarrow 0, k) \sim \left( \frac{k}{2 k_F}-1\right) \ln \left( \frac{k}{2 k_F}-1\right).
\end{equation}
We numerically evaluate $G_R(\omega, k) \sim \frac{\mathcal{B}}{\mathcal{A}}$ in this limit. Figure \ref{fig:ratio} shows a plot  of $G_R(0, \widetilde{k})$ for a wide range of $\widetilde{k}$ and various temperatures.
Of course, since \cite{Kulaxizi:2008jx} did not find any evidence for a sharp Fermi surface at zero temperature, we do not expect to find traces of a Fermi surface in the small frequency limit of the retarded Green's functions at non-zero temperature either, and merely intend to corroborate their result. Indeed, we do not observe any characteristic structure consistent with a singularity in the Green's functions. In \cite{Lifschytz:2009sz}, the authors explored if the model exhibits quantum oscillations associated with a series of crossings of the Landau levels with the Fermi surface. They concluded
that the model does not possess de Haas-van Alphen oscillations, even at zero temperature. Similarly, our analysis does not seem to lead to further evidence for an observable Fermi surface, despite the existence of a collective excitation corresponding to zero sound. Some possible implications are listed in the conclusions.

\section{Conclusions and outlook}

The principal line of research followed in the present work was to clarify the nature of holographic quantum liquids, in particular, their low energy effective description at finite density, and to compare them with properties of real materials such as liquid Helium-3, which at sufficiently low temperatures are well understood in terms of Landau's theory of Fermi liquids.

We focused on the deconfined phase of the Sakai-Sugimoto model of holographic QCD, expanding on previous results regarding the existence of a zero sound mode in the longitudinal channel of the current-current correlators. More specifically, following \cite{Bergman:2011rf,Davison:2011ek}, we generalized the zero-temperature (or infinite baryonic density) results of \cite{Kulaxizi:2008jx} for the zero sound mode to arbitrary values of $T$ and $\mu$ and we studied in detail its behavior as $T/\mu$ was increased.

Let us summarize some of the important findings of the Sakai-Sugimoto model at finite temperature and baryon density/chemical potential:
\begin{itemize}
\item Some thermodynamic properties of the deconfined phase of the Sakai-Sugimoto model appear to be consistent with a strongly-coupled Fermi liquid interpretation. We must emphasize that the existence of such a phase at high enough baryon density is well established in perturbative QCD \cite{Son:2000xc,Son:2000by,Evans:1998ek}, which means that the existence of a Fermi surface seems to naively extrapolate to the strong coupling regime. In particular, the specific heat at low temperatures scales like $C_v \sim T$  \cite{Kulaxizi:2008jx}, in agreement with the predictions of Landau-Fermi theory, while in the standard $D3/D7$ case $C_v \sim T^6$ \cite{Davison:2011ek}. On the other hand, the energy density varies as $n_B^{5/3}$, which is the expected equation of state for non-relativistic fermions \cite{Kim:2007vd}.
\item The model features a stable zero sound mode with a non-standard $k^3$ dependence of the attenuation rate at low and moderate temperatures (collisionless quantum and collisionless thermal regimes).
\item For large enough temperatures (hydrodynamic regime), the zero sound mode is effectively damped out by thermal fluctuations. The low energy physics in this case is dominated by an emergent diffusive mode, in agreement with the results of \cite{Kim:2008bv}.
\item The behavior of the attenuation rate at finite temperature in the three regimes discussed above is reminiscent of Landau-Fermi liquid theory, albeit with different temperature dependences. In particular, we found that $\Gamma_T\sim$ constant in the collisionless quantum regime and $\Gamma_T\sim T^3$ in the collisionless thermal regime. In the hydrodynamic regime, we can distinguish between a cold and a hot phase where $\Gamma_T\sim T^{-3}$ and $\Gamma_T\sim T^{-1}$, respectively. In the latter regime, the attenuation rate refers to the diffusion mode, while in the first two cases, to the zero sound mode. 
\item We did not discover any evidence for a direct signature of a sharp Fermi surface at zero or non-zero temperature. One possible explanation is that the Fermi surface structure expected in the spectral functions at low frequencies might only appear at order $N_f/N_c$, in which case we would need to include backreaction to have a visible effect.\footnote{Some backreacted configurations in the Sakai-Sugimoto model were studied in \cite{Burrington:2007qd}.} If this is the case, the first step would be to construct the backreacted background at finite chemical potential with the fluctuations turned off. On top of this background we would need to consider small fluctuations. Gauge fluctuations will generically mix with metric fluctuations and we would have to consider an appropriate gauge invariant combination for the sound channel as in \cite{Edalati:2010pn}. The second possibility is that the zero sound mode that we observe is not actually due to the presence of a Fermi surface, but rather, should be 
interpreted as a Goldstone boson arising from the spontaneous breaking of a specific symmetry, see e.g. \cite{Nickel:2010pr,Edalati:2013tma}. Both options would be interesting in their own right. Either we confirm the existence of a Fermi liquid phase at strong coupling or we discover a new low energy description of strongly-correlated electrons with no Fermi surfaces.
\item At large chemical potential to temperature ratio $\mu/T$ the model is expected to decay to a spatially inhomogeneous phase \cite{Ooguri:2010xs}. 
The critical value $\tilde\mu\sim 3.7$ for the on-set of the instability is 
of the same order as the threshold for the zero sound to be completely damped by thermal effects and ceasing to propagate, thus leaving only a fairly small range of parameter space for the zero sound. 
However, to precisely address this question one should first construct the striped solution out of the equations of motion following from the DBI+CS action, 
and compare the free energies to decipher which phase dominates.
\end{itemize}

Thus, in conclusion, although some of the features of the model are compatible with the Landau theory of Fermi liquids, further research is necessary in order to determine whether this model of holographic QCD has a true low energy description in terms of quasiparticle excitations of a Fermi surface or if it constitutes a novel type of strongly-coupled quantum liquid phase without a Fermi surface at all.

As mentioned in the introduction, the inclusion of a large magnetic field 
in the worldvolume of the flavor branes would be an interesting extension of the present work because such a case may not be unstable to the formation of a spatially modulated phase. 
The zero sound mode is expected to develop a gap in the dispersion relation, and would lead to interesting physics regarding, e.g, the fate of Kohn's theorem for nonrelativistic fermions with pairwise interactions. Another interesting possibility for future research would be to investigate the case of non-supersymmetric flavor $D7/\overline{D7}$- and $D5/\overline{D5}$-brane setups in the Klebanov-Witten and Klebanov-Strassler backgrounds \cite{Klebanov:1998hh, Kuperstein:2008cq, Dymarsky:2009cm, Bayona:2010bg, Ihl:2010zg,Alam:2012fw, Ihl:2012bm, Alam:2013cia, Filev:2013vka}. Both cases are currently under consideration.

\acknowledgments
The authors would like to acknowledge very useful conversations and correspondence with Alfonso Ballon Bayona, Richard Davison, Mohammad Edalati, Matthias Kaminski, Matthew Lippert, Alfonso Ramallo, Gordon Semenoff, and Dimitrios Zoakos. M.I. is grateful to the Theory Group of the Department of Physics at the University of Texas at Austin, the Department of Physics at the University of Washington, the Department of Physics at the University of British Columbia, and the Theory Group, Instituto Galego de F{\'i}sica de Altas Enerx{\'i}as, Universidade de Santiago de Compostela for hospitality during intermediate stages of this work; moreover, M.I. would like to express his gratitude to Estibalitz Ukar and Christoph Sachse for hospitality during his extended stay in Austin, TX. The work of B.D and J.P is partially supported by the National Science Foundation under grant Grant No. PHY-1316033 and by the Texas Cosmology Center. M.I. is funded by the FCT fellowship SFRH/BI/52188/2013. The Centro de F\'isica do 
Porto is partially funded by FCT through the projects PTDC/FIS/099293/2008 and CERN/FP/116358/2010.
N.J. is funded by the Spanish grant FPA2011-22594, by the Consolider-Ingenio 2010 Programme CPAN (CSD2007-00042), by Xunta de Galicia (GRC2013-024), and by FEDER. N.J. is also supported by the Juan de la Cierva program.


\begin{thebibliography}{99}

\bibitem{Maldacena:1997re}
J.~M.~Maldacena, ``The Large N limit of superconformal field theories and supergravity,'' Adv.\ Theor.\ Math.\ Phys.\  {\bf 2}, 231 (1998), Int.\ J.\ Theor.\ Phys.\  {\bf 38}, 1113 (1999), [hep-th/9711200].

\bibitem{Gubser:1998bc}
  S.~S.~Gubser, I.~R.~Klebanov and A.~M.~Polyakov,
  ``Gauge theory correlators from non-critical string theory,''
  Phys.\ Lett.\  B {\bf 428}, 105 (1998)
  [arXiv:hep-th/9802109].

\bibitem{Witten:1998qj}
  E.~Witten,
  ``Anti-de Sitter space and holography,''
  Adv.\ Theor.\ Math.\ Phys.\  {\bf 2}, 253 (1998)
  [arXiv:hep-th/9802150].

\bibitem{Sakai:2004cn}
  T.~Sakai and S.~Sugimoto,
  ``Low energy hadron physics in holographic QCD,''
  Prog.\ Theor.\ Phys.\  {\bf 113}, 843 (2005)
  [hep-th/0412141].

\bibitem{Sakai:2005yt}
  T.~Sakai and S.~Sugimoto,
  ``More on a holographic dual of QCD,''
  Prog.\ Theor.\ Phys.\  {\bf 114}, 1083 (2005)
  [hep-th/0507073].

\bibitem{Kim:2008bv}
  K.~-Y.~Kim and I.~Zahed,
  ``Baryonic Response of Dense Holographic QCD,''
  JHEP {\bf 0812}, 075 (2008)
  [arXiv:0811.0184 [hep-th]].

\bibitem{Kulaxizi:2008jx}
  M.~Kulaxizi and A.~Parnachev,
  ``Holographic Responses of Fermion Matter,''
  Nucl.\ Phys.\ B {\bf 815}, 125 (2009)
  [arXiv:0811.2262 [hep-th]].

\bibitem{Kim:2007vd}
  K.~-Y.~Kim, S.~-J.~Sin and I.~Zahed,
  ``Dense holographic QCD in the Wigner-Seitz approximation,''
  JHEP {\bf 0809}, 001 (2008)
  [arXiv:0712.1582 [hep-th]].

\bibitem{Bergman:2007wp}
  O.~Bergman, G.~Lifschytz and M.~Lippert,
  ``Holographic Nuclear Physics,''
  JHEP {\bf 0711} (2007) 056
  [arXiv:0708.0326 [hep-th]].


\bibitem{Rozali:2007rx}
  M.~Rozali, H.~-H.~Shieh, M.~Van Raamsdonk and J.~Wu,
  ``Cold Nuclear Matter In Holographic QCD,''
  JHEP {\bf 0801}, 053 (2008)
  [arXiv:0708.1322 [hep-th]].

\bibitem{Son:2000xc}
  D.~T.~Son and M.~A.~Stephanov,
  ``QCD at finite isospin density,''
  Phys.\ Rev.\ Lett.\  {\bf 86}, 592 (2001)
  [hep-ph/0005225].

\bibitem{Son:2000by}
  D.~T.~Son and M.~A.~Stephanov,
  ``QCD at finite isospin density: From pion to quark - anti-quark condensation,''
  Phys.\ Atom.\ Nucl.\  {\bf 64}, 834 (2001)
  [Yad.\ Fiz.\  {\bf 64}, 899 (2001)]
  [hep-ph/0011365].

\bibitem{Evans:1998ek}
  N.~J.~Evans, S.~D.~H.~Hsu and M.~Schwetz,
  ``An Effective field theory approach to color superconductivity at high quark density,''
  Nucl.\ Phys.\ B {\bf 551}, 275 (1999)
  [hep-ph/9808444].

\bibitem{Landau}
 L.~D.~Landau,
 ``Oscillations in a Fermi liquid,''
 Zh.\ Eksp.\ Teor.\ Fiz.\ {\bf 32}, 59 (1957)
 [Soviet Phys. - JETP {\bf 5}, 101 (1959)].

\bibitem{Karch:2008fa}
  A.~Karch, D.~T.~Son and A.~O.~Starinets,
  ``Zero Sound from Holography,''
  arXiv:0806.3796 [hep-th].

  \bibitem{Bergman:2011rf}
  O.~Bergman, N.~Jokela, G.~Lifschytz and M.~Lippert,
  ``Striped instability of a holographic Fermi-like liquid,''
  JHEP {\bf 1110}, 034 (2011)
  [arXiv:1106.3883 [hep-th]].

  \bibitem{Jokela:2012vn}
  N.~Jokela, G.~Lifschytz and M.~Lippert,
  ``Magnetic effects in a holographic Fermi-like liquid,''
  JHEP {\bf 1205}, 105 (2012)
  [arXiv:1204.3914 [hep-th]].

\bibitem{Kulaxizi:2008kv}
  M.~Kulaxizi and A.~Parnachev,
  ``Comments on Fermi Liquid from Holography,''
  Phys.\ Rev.\ D {\bf 78}, 086004 (2008)
  [arXiv:0808.3953 [hep-th]].

\bibitem{Hung:2009qk}
  L.~-Y.~Hung and A.~Sinha,
  ``Holographic quantum liquids in 1+1 dimensions,''
  JHEP {\bf 1001}, 114 (2010)
  [arXiv:0909.3526 [hep-th]].
  
\bibitem{Edalati:2010pn}
  M.~Edalati, J.~I.~Jottar and R.~G.~Leigh,
  ``Holography and the sound of criticality,''
  JHEP {\bf 1010}, 058 (2010)
  [arXiv:1005.4075 [hep-th]].
  
\bibitem{HoyosBadajoz:2010kd}
  C.~Hoyos-Badajoz, A.~O'Bannon and J.~M.~S.~Wu,
  ``Zero Sound in Strange Metallic Holography,''
  JHEP {\bf 1009}, 086 (2010)
  [arXiv:1007.0590 [hep-th]].

\bibitem{Nickel:2010pr}
  D.~Nickel and D.~T.~Son,
  ``Deconstructing holographic liquids,''
  New J.\ Phys.\  {\bf 13}, 075010 (2011)
  [arXiv:1009.3094 [hep-th]].

\bibitem{Lee:2010ez}
  B.~-H.~Lee, D.~-W.~Pang and C.~Park,
  ``Zero Sound in Effective Holographic Theories,''
  JHEP {\bf 1011}, 120 (2010)
  [arXiv:1009.3966 [hep-th]].

  
\bibitem{Ammon:2011hz}
  M.~Ammon, J.~Erdmenger, S.~Lin, S.~Muller, A.~O'Bannon, J.~P.~Shock, J.~Erdmenger and S.~Lin {\it et al.},
  ``On Stability and Transport of Cold Holographic Matter,''
  JHEP {\bf 1109}, 030 (2011)
  [arXiv:1108.1798 [hep-th]].

 \bibitem{Davison:2011ek}
  R.~A.~Davison and A.~O.~Starinets,
  ``Holographic zero sound at finite temperature,''
  Phys.\ Rev.\ D {\bf 85}, 026004 (2012)
  [arXiv:1109.6343 [hep-th]].

\bibitem{Davison:2011uk}
  R.~A.~Davison and N.~K.~Kaplis,
  ``Bosonic excitations of the $AdS_4$ Reissner-Nordstrom black hole,''
  JHEP {\bf 1112}, 037 (2011)
  [arXiv:1111.0660 [hep-th]].


\bibitem{Goykhman:2012vy}
  M.~Goykhman, A.~Parnachev and J.~Zaanen,
  ``Fluctuations in finite density holographic quantum liquids,''
  JHEP {\bf 1210}, 045 (2012)
  [arXiv:1204.6232 [hep-th]].

\bibitem{Gorsky:2012gi}
  A.~Gorsky and A.~V.~Zayakin,
  ``Anomalous Zero Sound,''
  JHEP {\bf 1302}, 124 (2013)
  [arXiv:1206.4725 [hep-th]].

\bibitem{Brattan:2012nb}
  D.~K.~Brattan, R.~A.~Davison, S.~A.~Gentle and A.~O'Bannon,
  ``Collective Excitations of Holographic Quantum Liquids in a Magnetic Field,''
  JHEP {\bf 1211}, 084 (2012)
  [arXiv:1209.0009 [hep-th]].

\bibitem{Jokela:2012se}
  N.~Jokela, M.~Jarvinen and M.~Lippert,
  ``Fluctuations and instabilities of a holographic metal,''
  JHEP {\bf 1302}, 007 (2013)
  [arXiv:1211.1381 [hep-th]].

\bibitem{Pang:2013ypa}
  D.~-W.~Pang,
  ``Probing holographic semi-local quantum liquids with D-branes,''
  Phys.\ Rev.\ D {\bf 88}, 046002 (2013)
  [arXiv:1306.3816 [hep-th]].

\bibitem{Dey:2013vja}
  P.~Dey and S.~Roy,
  ``Zero sound in strange metals with hyperscaling violation from holography,''
  Phys.\ Rev.\ D {\bf 88}, 046010 (2013)
  [arXiv:1307.0195 [hep-th]].

\bibitem{Edalati:2013tma}
  M.~Edalati and J.~F.~Pedraza,
  ``Aspects of Current Correlators in Holographic Theories with Hyperscaling Violation,''
  Phys.\ Rev.\ D {\bf 88}, 086004 (2013)
  [arXiv:1307.0808 [hep-th]].

\bibitem{Davison:2013uha}
  R.~A.~Davison, M.~Goykhman and A.~Parnachev,
  ``AdS/CFT and Landau Fermi liquids,''
  arXiv:1312.0463 [hep-th].


\bibitem{Rozali:2014yva}
  M.~Rozali and D.~Smyth,
  ``Fermi Liquids from D-Branes,''
  arXiv:1402.7043 [hep-th].

\bibitem{Liu:2009dm}
  H.~Liu, J.~McGreevy and D.~Vegh,
  ``Non-Fermi liquids from holography,''
  Phys.\ Rev.\ D {\bf 83}, 065029 (2011)
  [arXiv:0903.2477 [hep-th]].

\bibitem{Iqbal:2011ae}
  N.~Iqbal, H.~Liu and M.~Mezei,
  ``Lectures on holographic non-Fermi liquids and quantum phase transitions,''
  arXiv:1110.3814 [hep-th].

\bibitem{Jensen:2011su}
  K.~Jensen, S.~Kachru, A.~Karch, J.~Polchinski and E.~Silverstein,
  ``Towards a holographic marginal Fermi liquid,''
  Phys.\ Rev.\ D {\bf 84}, 126002 (2011)
  [arXiv:1105.1772 [hep-th]].

\bibitem{Ooguri:2010xs}
  H.~Ooguri and C.~-S.~Park,
  ``Spatially Modulated Phase in Holographic Quark-Gluon Plasma,''
  Phys.\ Rev.\ Lett.\  {\bf 106} (2011) 061601
  [arXiv:1011.4144 [hep-th]].

\bibitem{Bergman:2010gm}
  O.~Bergman, N.~Jokela, G.~Lifschytz and M.~Lippert,
  ``Quantum Hall Effect in a Holographic Model,''
  JHEP {\bf 1010} (2010) 063
  [arXiv:1003.4965 [hep-th]].

\bibitem{Jokela:2011eb}
  N.~Jokela, M.~Jarvinen and M.~Lippert,
  ``A holographic quantum Hall model at integer filling,''
  JHEP {\bf 1105} (2011) 101
  [arXiv:1101.3329 [hep-th]].

\bibitem{Lifschytz:2009sz}
  G.~Lifschytz and M.~Lippert,
  ``Holographic Magnetic Phase Transition,''
  Phys.\ Rev.\ D {\bf 80} (2009) 066007
  [arXiv:0906.3892 [hep-th]].

\bibitem{Preis:2010cq}
  F.~Preis, A.~Rebhan and A.~Schmitt,
  ``Inverse magnetic catalysis in dense holographic matter,''
  JHEP {\bf 1103} (2011) 033
  [arXiv:1012.4785 [hep-th]].

\bibitem{Aharony:2007uu}
  O.~Aharony, K.~Peeters, J.~Sonnenschein and M.~Zamaklar,
  ``Rho meson condensation at finite isospin chemical potential in a holographic model for QCD,''
  JHEP {\bf 0802} (2008) 071
  [arXiv:0709.3948 [hep-th]].

\bibitem{Bayona:2011ab}
  C.~A.~B.~Bayona, K.~Peeters and M.~Zamaklar,
  ``A Non-homogeneous ground state of the low-temperature Sakai-Sugimoto model,''
  JHEP {\bf 1106} (2011) 092
  [arXiv:1104.2291 [hep-th]].

\bibitem{BallonBayona:2012wx}
  A.~Ballon-Bayona, K.~Peeters and M.~Zamaklar,
  ``A chiral magnetic spiral in the holographic Sakai-Sugimoto model,''
  JHEP {\bf 1211} (2012) 164
  [arXiv:1209.1953 [hep-th]].

\bibitem{Aharony:2006da}
  O.~Aharony, J.~Sonnenschein and S.~Yankielowicz,
  ``A Holographic model of deconfinement and chiral symmetry restoration,''
  Annals Phys.\  {\bf 322}, 1420 (2007)
  [hep-th/0604161].

\bibitem{Mandal:2011ws} 
  G.~Mandal and T.~Morita,
  ``Gregory-Laflamme as the confinement/deconfinement transition in holographic QCD,''
  JHEP {\bf 1109}, 073 (2011)
  [arXiv:1107.4048 [hep-th]].

\bibitem{Mandal:2011uq} 
  G.~Mandal and T.~Morita,
  ``What is the gravity dual of the confinement/deconfinement transition in holographic QCD?,''
  J.\ Phys.\ Conf.\ Ser.\  {\bf 343}, 012079 (2012)
  [arXiv:1111.5190 [hep-th]].


\bibitem{Kobayashi:2006sb}
  S.~Kobayashi, D.~Mateos, S.~Matsuura, R.~C.~Myers and R.~M.~Thomson,
  ``Holographic phase transitions at finite baryon density,''
  JHEP {\bf 0702}, 016 (2007)
  [hep-th/0611099].

\bibitem{Kim:2007zm}
  K.~-Y.~Kim, S.~-J.~Sin and I.~Zahed,
  ``The Chiral Model of Sakai-Sugimoto at Finite Baryon Density,''
  JHEP {\bf 0801} (2008) 002
  [arXiv:0708.1469 [hep-th]].

  
\bibitem{Kim:2006gp}
  K.~-Y.~Kim, S.~-J.~Sin and I.~Zahed,
  ``Dense hadronic matter in holographic QCD,''
  J.\ Korean Phys.\ Soc.\  {\bf 63} (2013) 1515
  [hep-th/0608046].
  
\bibitem{Horigome:2006xu}
  N.~Horigome and Y.~Tanii,
  ``Holographic chiral phase transition with chemical potential,''
  JHEP {\bf 0701} (2007) 072
  [hep-th/0608198].
  
\bibitem{Sin:2007ze}
  S.~-J.~Sin,
  ``Gravity back-reaction to the baryon density for bulk filling branes,''
  JHEP {\bf 0710} (2007) 078
  [arXiv:0707.2719 [hep-th]].

\bibitem{Yamada:2007ys}
  D.~Yamada,
  ``Sakai-Sugimoto model at high density,''
  JHEP {\bf 0810} (2008) 020
  [arXiv:0707.0101 [hep-th]].




  \bibitem{Parnachev:2006ev}
  A.~Parnachev and D.~A.~Sahakyan,
  ``Photoemission with Chemical Potential from QCD Gravity Dual,''
  Nucl.\ Phys.\ B {\bf 768}, 177 (2007)
  [hep-th/0610247].

\bibitem{Son:2002sd}
  D.~T.~Son and A.~O.~Starinets,
  ``Minkowski space correlators in AdS / CFT correspondence: Recipe and applications,''
  JHEP {\bf 0209}, 042 (2002)
  [hep-th/0205051].


  \bibitem{Herzog:2002pc}
  C.~P.~Herzog and D.~T.~Son,
  ``Schwinger-Keldysh propagators from AdS/CFT correspondence,''
  JHEP {\bf 0303}, 046 (2003)
  [hep-th/0212072].


  \bibitem{CaronHuot:2006te}
  S.~Caron-Huot, P.~Kovtun, G.~D.~Moore, A.~Starinets and L.~G.~Yaffe,
  ``Photon and dilepton production in supersymmetric Yang-Mills plasma,''
  JHEP {\bf 0612}, 015 (2006)
  [hep-th/0607237].


 \bibitem{Kaminski:2009dh}
  M.~Kaminski, K.~Landsteiner, J.~Mas, J.~P.~Shock and J.~Tarrio,
  ``Holographic Operator Mixing and Quasinormal Modes on the Brane,''
  JHEP {\bf 1002}, 021 (2010)
  [arXiv:0911.3610 [hep-th]].

\bibitem{Burrington:2007qd}
  B.~A.~Burrington, V.~S.~Kaplunovsky and J.~Sonnenschein,
  ``Localized Backreacted Flavor Branes in Holographic QCD,''
  JHEP {\bf 0802}, 001 (2008)
  [arXiv:0708.1234 [hep-th]].

\bibitem{Klebanov:1998hh}
  I.~R.~Klebanov and E.~Witten,
  ``Superconformal field theory on three-branes at a Calabi-Yau singularity,''
  Nucl.\ Phys.\ B {\bf 536}, 199 (1998)
  [hep-th/9807080].

\bibitem{Kuperstein:2008cq}
  S.~Kuperstein and J.~Sonnenschein,
  ``A New Holographic Model of Chiral Symmetry Breaking,''
  JHEP {\bf 0809}, 012 (2008)
  [arXiv:0807.2897 [hep-th]].

  \bibitem{Dymarsky:2009cm}
  A.~Dymarsky, S.~Kuperstein and J.~Sonnenschein,
  ``Chiral Symmetry Breaking with non-SUSY D7-branes in ISD backgrounds,''
  JHEP {\bf 0908}, 005 (2009)
  [arXiv:0904.0988 [hep-th]].

\bibitem{Bayona:2010bg}
  C.~A.~B.~Bayona, H.~Boschi-Filho, M.~Ihl and M.~A.~C.~Torres,
  ``Pion and Vector Meson Form Factors in the Kuperstein-Sonnenschein holographic model,''
  JHEP {\bf 1008}, 122 (2010)
  [arXiv:1006.2363 [hep-th]].

\bibitem{Ihl:2010zg}
  M.~Ihl, M.~A.~C.~Torres, H.~Boschi-Filho and C.~A.~B.~Bayona,
  ``Scalar and vector mesons of flavor chiral symmetry breaking in the Klebanov-Strassler background,''
  JHEP {\bf 1109}, 026 (2011)
  [arXiv:1010.0993 [hep-th]].

\bibitem{Alam:2012fw}
  M.~S.~Alam, V.~S.~Kaplunovsky and A.~Kundu,
  ``Chiral Symmetry Breaking and External Fields in the Kuperstein-Sonnenschein Model,''
  JHEP {\bf 1204}, 111 (2012)
  [arXiv:1202.3488 [hep-th]].

\bibitem{Ihl:2012bm}
  M.~Ihl, A.~Kundu and S.~Kundu,
  ``Back-reaction of Non-supersymmetric Probes: Phase Transition and Stability,''
  JHEP {\bf 1212}, 070 (2012)
  [arXiv:1208.2663 [hep-th]].

\bibitem{Alam:2013cia}
  M.~Sohaib Alam, M.~Ihl, A.~Kundu and S.~Kundu,
  ``Dynamics of Non-supersymmetric Flavours,''
  JHEP {\bf 1309}, 130 (2013)
  [arXiv:1306.2178 [hep-th]].

 \bibitem{Filev:2013vka}
  V.~G.~Filev, M.~Ihl and D.~Zoakos,
  ``A Novel (2+1)-Dimensional Model of Chiral Symmetry Breaking,''
  JHEP {\bf 1312}, 072 (2013)
  [arXiv:1310.1222 [hep-th]].

\end{thebibliography}
\end{document}